\documentclass[lettersize,onecolumn]{IEEEtran}
\usepackage{amsmath,amsfonts}
\usepackage{amssymb}
\usepackage{algorithmic}
\usepackage{algorithm}
\usepackage{array}
\usepackage[caption=false,font=normalsize,labelfont=sf,textfont=sf]{subfig}
\usepackage{textcomp}
\usepackage{stfloats}
\usepackage{url}
\usepackage{verbatim}
\usepackage{graphicx}
\usepackage{cite}
\usepackage{color}
\usepackage{amsthm}
\newtheorem{Theorem}{Theorem}
\newtheorem{Proposition}{Proposition}

\newtheorem{Lemma}{Lemma}
\newtheorem{Corollary}{Corollary}
\newtheorem{Remark}{Remark}

\usepackage[dvipsnames]{xcolor}
\usepackage{hyperref}
\hypersetup{colorlinks,linkcolor={BrickRed},citecolor={blue},urlcolor={blue}}

\usepackage{enumitem}
\allowdisplaybreaks

\begin{document}

\title{Soft Guessing Under Logarithmic Loss Allowing Errors and Variable-Length Source Coding}

\author{Shota~Saito~and~Hamdi~Joudeh
\thanks{This paper was presented in part at the 2024 IEEE International Symposium on Information Theory \cite{SaitoISIT24}, the 2024 International Symposium on Information Theory and its Applications \cite{SaitoISITA24}, and the 47th Symposium on Information Theory and its Applications \cite{SaitoSITA24}. This work was supported in part by JSPS KAKENHI Grant Numbers JP23K11097, JP23H00468, and JP25K07732, and in part by the European Research Council (ERC) under Grant Number 101116550.}
\thanks{Shota Saito is with the Faculty of Informatics, Gunma University, 4-2, Maebashi, Gunma 371-8510, Japan (e-mail: shota.s@gunma-u.ac.jp). Hamdi Joudeh is with the Department of Electrical Engineering, Eindhoven University of Technology, 5600 MB Eindhoven, The Netherlands (e-mail: h.joudeh@tue.nl).}
}

\maketitle

\begin{abstract}
This paper considers the problem of soft guessing under a logarithmic loss distortion measure while allowing errors. We find an optimal guessing strategy, and derive single-shot upper and lower bounds for the minimal guessing moments as well as an asymptotic expansion for i.i.d. sources. These results are extended to the case where side information is available to the guesser. Furthermore, a connection between soft guessing allowing errors and variable-length lossy source coding under logarithmic loss is demonstrated. The R\'enyi entropy, the smooth R\'enyi entropy, and their conditional versions play an important role.
\end{abstract}

\begin{IEEEkeywords}
Guessing, log-loss, R\'enyi entropy, smooth R\'enyi entropy, variable-length lossy source coding.
\end{IEEEkeywords}

\section{Introduction}

%\subsection{Soft Guessing Under log-loss}
%\IEEEPARstart{I}{n} information-theoretic literature, the problem of guessing is one of the research topics. 
\IEEEPARstart{I}{n} 1994, Massey \cite{Massey} introduced the information-theoretic study of guessing, or guesswork, and considered the following basic problem: a guesser seeks to determine the value of a random variable $X$, taking values in $\mathcal{X}=\{1, 2, \ldots, |\mathcal{X}| \}$, by asking questions of the form: ``is $X$ equal to $x_1$?'', ``is $X$ equal to $x_2$?'' and so on. An honest answer is returned for each query, and guessing continues until the answer is ``yes''.
%%%%%
Such a guessing strategy is specified by a permutation of $\mathcal{X}$, and is denoted henceforth as $\mathcal{G} = (x_1,x_2,\ldots,x_{|\mathcal{X}|})$.
%%%%%%
Any guessing strategy induces a bijective function $g: \mathcal{X} \to \{1,\ldots,|\mathcal{X}|\}$, 
where $g(x)$ is the guessing order of $x$, i.e., the number of guesses required when $X = x$.
%%%
Massey \cite{Massey} investigated the relationship between the expected number of guesses, minimized over all guessing strategies, and the Shannon entropy.

Two years later, Ar\i kan \cite{Arikan} considered the more general problem of studying the minimal $\rho$-th guessing moment,  defined for every  $\rho > 0$ as
\begin{align}
    M^\star_X (\rho) := \min_{\mathcal{G}} \mathbb{E} \left[ g(X)^\rho \right] 
    = \min_{\mathcal{G}} \sum_{x \in \mathcal{X}} P_X(x) g(x)^\rho,
\end{align}
where  $P_X$ is the probability mass function (pmf) of $X$. Ar\i kan showed that all guessing moments are simultaneously minimized by a strategy that queries realizations in a decreasing order of their probability, and derived bounds on $M^\star_X (\rho)$ given by
%%%%%%
\begin{equation}
    M^\star_X (\rho) \leq \exp\left( \rho H_{\frac{1}{1+\rho}}(X) \right) 
    \label{eq:upper_bound_Arikan}
\end{equation}
and
\begin{equation}
   M^\star_X (\rho) \geq  \left(1 + \log |\mathcal{X}| \right)^{-\rho} \exp\left( \rho H_{\frac{1}{1+\rho}}(X) \right),
  \label{eq:lower_bound_Arikan}
\end{equation}
%%%
%%%%%
where $H_{\frac{1}{1+\rho}}(X)$ is the R\'enyi entropy of order $1/(1+\rho)$ \cite{Renyi}  (see Section \ref{Definition_RE}).
%%%%%
Together, \eqref{eq:upper_bound_Arikan} and \eqref{eq:lower_bound_Arikan} provide a tight exponential characterization of the guessing moment (known as the guessing exponent) in the i.i.d. asymptotic regime, where the task is to guess an $n$-vector $X^n = X_1,\ldots,X_n$ of i.i.d. entries.
%%%%%
%In particular, using the fact that $H_{\frac{1}{1+\rho}} (X^n) = nH_{\frac{1}{1+\rho}} (X)$, it follows that 
%\begin{equation}
%    E_X(\rho) := \lim_{n \to \infty} \frac{1}{n} \log M_{X^n}^{\star}(\rho) = \rho H_{\frac{1}{1+\rho}} (X)
%\end{equation}
%%%%%
%where $E_X(\rho)$ is known as the $\rho$-th guessing exponent. 
Ar\i kan further generalized the problem to the case where a correlated side information random variable is available to the guesser, establishing a connection to the Arimoto-R\'enyi conditional entropy \cite{Arimoto}; and applied this result to study the computational complexity and cutoff rate of sequential decoding.

Since the works of Massey and Ar\i kan, the original guessing problem has been extended and studied in various contexts, including: guessing subject to distortion \cite{ArikanMerhav,Cohen,Merhav1999,Saito,Wu}, guessing allowing errors \cite{Kuzuoka,Sakai}, guessing under source uncertainty \cite{Sundaresan}, guessing and large deviations \cite{Arikan2000, Christiansen, Hanawal, Pfister, SundaresanISIT}, guessing and joint source-channel coding \cite{ArikanMerhav2}, guessing via an unreliable oracle \cite{Burin}, guessing with limited memory \cite{Huleihel}, guessing for Markov sources \cite{Malone}, multi-agent guesswork \cite{Salamatian, Salamatian2}, guesswork of hash functions \cite{Yona}, multi-user guesswork \cite{Christiansen2}, guesswork subject to a per-symbol Shannon entropy budget \cite{Beirami}, universal randomized guessing \cite{Cohen, Merhav}, guessing based on compressed side information \cite{Graczyk}, guessing individual sequences \cite{Merhav2020}, guesses transmitted via a noisy channel \cite{MerhavNoisyGuess}, multiple guesses under a tunable loss function \cite{Kurri}, improved bounds and connections to variable-length source coding \cite{SasonVerdu}, and connections to majorization theory \cite{Sason}, among others.
%%%%%
Guessing has also played a key role in analyzing a channel decoding paradigm that queries noise patterns instead of codewords, known as GRAND \cite{Duffy2019,Duffy2022,Joudeh2024,Tan2025}.

Of particular interest to our current work are two extensions of the original Massey-Ar\i kan guessing settings: the first is the \emph{guessing allowing errors} problem proposed by Kuzuoka \cite{Kuzuoka}, and the second is the \emph{soft guessing under log-loss} problem proposed by Wu and Joudeh \cite{Wu}.
%%%%%
Next, we briefly review these two problems. 
%%%%%
\subsection{Guessing Allowing Errors}
\label{sec:guessing_allowing_errors_kuzuoka}
%%%%%
In Kuzuoka's guessing allowing errors framework, the guesser may stop guessing and declare an error at any step \cite{Kuzuoka}. In particular, at the $i$-th step, the guesser gives up guessing and declares an error with probability $\pi_i$ or continues guessing with probability $1-\pi_i$, where $0 \leq \pi_i \leq 1$. A guessing strategy is specified by a pair $(\mathcal{G},\pi)$, where $\mathcal{G}$ is as defined earlier, inducing a guessing function $g$, while $\pi= (\pi_1,\pi_2,\ldots,\pi_{|\mathcal{X}|})$ is a sequence of give up probabilities.
For a fixed strategy, let 
\begin{align}
\label{eq:lambda_i}
    \lambda_i := \prod_{j=1}^{i} (1-\pi_j)
\end{align}
for $i=1, 2, \ldots, |\mathcal{X}|$, i.e., $\lambda_i$ is the probability that the guesser does not give up guessing before making the $i$-th guess.
%%%%%
Such a randomized guessing strategy gives rise to a stochastic guessing function $G: \mathcal{X} \to \{1,\ldots,|\mathcal{X}|\} \cup \{0\}$ defined as
\begin{align}
\label{eq:stochastic_G}
G(x) := 
\begin{cases}
g(x), & \mathrm{with ~ probability}~ \lambda_{g(x)}, \\
0, & \mathrm{with ~ probability}~ 1-\lambda_{g(x)}.
\end{cases}
\end{align}
%%%%%
Note that $G(x) = 0$ represents the event that the guesser gives up before asking the question ``is $X$ equal to $x$?''.
%%%%%%
The probability that $X$ is correctly guessed at the $i$-th step before giving up is hence  $\mathbb{P}\left[G(X) = i \right] = \lambda_i P_X \big(g^{-1}(i) \big)$. On the other hand, the probability that the guesser gives up before correctly guessing $X$, i.e., error probability, is given by 
\begin{equation}
    P_e := \mathbb{P}\left[G(X) = 0 \right]
    = 1 - \sum_{i=1}^{|\mathcal{X}|} \lambda_i P_X \big(g^{-1}(i) \big).
\end{equation}
Therefore, the $\rho$-th guessing moment in this case is defined as
%%%%%%
\begin{equation}
M_X(\rho,\mathcal{G},\pi) := \mathbb{E}\left[G(X)^{\rho} \right] = \sum_{i=1}^{|\mathcal{X}|}  \lambda_i P_X \big(g^{-1}(i) \big) \times i^{\rho}.
\end{equation}
%%%%%%
Naturally, there is a trade-off between $M_X(\rho,\mathcal{G},\pi) $ and $P_e$, e.g., the former can be made arbitrarily small by making $P_e$ large enough.
%%%%%
Under the constraint that $P_e$ must not exceed $\epsilon$, the minimal $\rho$-th guessing moment is defined as\footnote{Kuzuoka also formulated the equivalent problem of minimizing the weighted sum $M_X(\rho,\mathcal{G},\pi) + \omega P_e$, where $\omega\geq 0$, but only solved \eqref{eq:guessing_moment_errors}. The formulation we adopt, with the stochastic guessing function as defined in \eqref{eq:stochastic_G}, is due to Sakai and Tan \cite[Section IV-C]{Sakai}.} 
\begin{equation}
\label{eq:guessing_moment_errors}
  M_X^{\star}(\rho,\epsilon) := \min_{\mathcal{G},\pi: P_e \leq \epsilon} M_X(\rho,\mathcal{G},\pi). 
\end{equation}
%%%%%
Mirroring  Ar\i kan's bounds in \eqref{eq:upper_bound_Arikan} and \eqref{eq:lower_bound_Arikan}, Kuzuoka \cite{Kuzuoka} showed that $  M_X^{\star}(\rho,\epsilon) $ is bounded as
%%%
\begin{equation}
    M^\star_X (\rho,\epsilon) \leq \exp\left( \rho H_{\frac{1}{1+\rho}}^{\epsilon}(X) \right) 
    \label{eq:upper_bound_Kuzuoka}
\end{equation}
%%%%
and
%%%%%%
\begin{equation}
   M^\star_X (\rho,\epsilon) \geq  \left(1 + \log |\mathcal{X}| \right)^{-\rho} \exp\left( \rho H_{\frac{1}{1+\rho}}^{\epsilon}(X) \right),
   \label{eq:lower_bound_Kuzuoka}
\end{equation}
%%%
where $H_{\frac{1}{1+\rho}}^{\epsilon}(X)$ denotes the $\epsilon$-smooth R\'enyi entropy, introduced by Renner and Wolf \cite{Renner05} (see Section \ref{Definition_SRE}). 
%%%%%
The bounds in \eqref{eq:upper_bound_Kuzuoka} and \eqref{eq:lower_bound_Kuzuoka} provide a tight exponential characterization of guessing moments in the i.i.d. asymptotic regime.
%, given as
%\begin{equation}
%    E_X(\rho,\epsilon) := \lim_{n \to \infty} \frac{1}{n} \log M_{X^n}^{\star}(\rho,\epsilon) = \rho H(X)
%\end{equation}
%%%%%
%where $H(X)$ denotes the Shannon entropy.
%%%%%
Interestingly, the guessing exponent %$E_X(\rho,\epsilon)$ 
in this case is characterized in terms of the Shannon entropy for any $\epsilon \in (0,1)$,  a consequence of the smooth R\'enyi entropy's asymptotic properties \cite{Sakai}.
%%%%%%
Kuzuoka also proposed an $\epsilon$-smooth counterpart to the  Arimoto-R\'enyi conditional entropy \cite{Kuzuoka}, and employed it to bound the $\rho$-th guessing moment for the side information variant of the problem.
%%%%%
\subsection{Soft Guessing Under Log-Loss}
\label{sec:soft_guessing_wu_joudeh}
%%%%%
The Wu-Joudeh soft guessing paradigm \cite{Wu} can be seen as a variant of guessing subject to distortion, but instead of guessing a ``hard'' reconstruction, i.e., a reproduction symbol $\hat{X}$ as in previous works (e.g., \cite{ArikanMerhav}), the guesser seeks to find a good ``soft'' reconstruction of $X$, i.e., a probability distribution $\hat{P}$ on the alphabet $\mathcal{X}$.
%%%%
The fidelity of the soft reconstruction is measured by the \emph{logarithmic loss} (log-loss) \cite{CourtadeWesel,CourtadeWeissman,Shkel}, defined for every symbol $x \in \mathcal{X}$
and soft reconstruction $\hat{P}$ as 
\begin{equation}
    d(x,\hat{P})  := \log \frac{1}{\hat{P}(x)}.
    \label{Def_LogLoss}
\end{equation}
%%%%%
A soft guessing strategy is specified by a sequence  $\mathcal{G}_{\mathrm{soft}}=(\hat{P}_1, \hat{P}_2, \ldots, \hat{P}_N)$ of $N$ probability distributions on $\mathcal{X}$, for some integer $N$. The guesser asks questions of the form: ``is $d(X, \hat{P}_1) \leq D$?'', ``is $d(X, \hat{P}_2) \leq D$?'' and so on until the answer is ``yes'', where $D \geq 0$ is some predetermined distortion level. 
%%%%%
If the soft guessing strategy terminates with probability $1$ for a given distortion level $D$, then we call it $D$-admissible, and we denote it by $\mathcal{G}_{\mathrm{soft}}(D)$.
%%%%%
Such a strategy induces a guessing function $g(x)$, which is the smallest index $j$ for which $d(x,\hat{P}_j) \leq D$ is satisfied.
%%%%%
Note that soft guessing can be reduced to standard guessing by setting $D = 0$ and selecting a strategy that consists of distinct hard reconstructions, i.e., single-mass point distributions.
%%%%%%
For a given distortion level $D$, the minimal $\rho$-the soft guessing moment is defined as 
\begin{align}
\label{eq:guessing_moment_soft}
    M^\star_X (\rho,D) := \min_{\mathcal{G}_{\mathrm{soft}}(D)} \mathbb{E} \left[ g(X)^\rho \right].
\end{align}
%%%%%%
Wu and Joudeh \cite{Wu} showed that $M^\star_X (\rho,D) $ is bounded as 
%%%
\begin{equation}
    M^\star_X (\rho,D) \leq 1 + 2^{\rho} \exp\left( \rho H_{\frac{1}{1+\rho}}(X) - \rho \log \lfloor \exp(D) \rfloor  \right) 
    \label{eq:upper_bound_wu_joudeh}
\end{equation}
%%%%
and
%%%%%%
\begin{equation}
   M^\star_X (\rho,D) \geq  \left(1 + \log |\mathcal{X}| \right)^{-\rho} \exp \left( \rho H_{\frac{1}{1+\rho}}(X) - \rho \log \lfloor \exp(D) \rfloor \right),
   \label{eq:lower_bound_wu_joudeh}
\end{equation}
%%%%
where the upper bound in \eqref{eq:upper_bound_wu_joudeh} can be tightened to Ar\i kan's upper bound in  \eqref{eq:upper_bound_Arikan} whenever $0 \leq D < 1$ (i.e. $\lfloor \exp(D) \rfloor = 1$).
%%%%%
A tight characterization of the guessing exponent in the i.i.d. asymptotic regime can also be recovered from \eqref{eq:upper_bound_wu_joudeh} and \eqref{eq:lower_bound_wu_joudeh}, and the result is also easily extended to the case where side information is available to the guesser.
%%%%%
\subsection{Soft Guessing Under Log-Loss Allowing Errors}
%%%%%
In this paper, we propose a natural generalization of the above settings. Specifically, we study the problem of soft guessing under log-loss while allowing errors.
%%%%%%
In this formulation, the guesser seeks a good soft reconstruction of $X$ in the sense of Wu and Joudeh \cite{Wu}, while also being permitted to give up guessing and declare an error, following the framework of Kuzuoka \cite{Kuzuoka}.
%%%%%%
A formal description of the setting is given in Section \ref{SectionSoftGuessingAlloingErrors}.
%%%%%%
Our goal is to derive upper and lower bounds on the minimal guessing moments in this setting, which subsume all previously stated bounds as special cases.

%%%%%%
Guessing subject to distortion is generally motivated by applications such as betting games, pattern matching, search algorithms, biometric authentication, and simple sequential rate–distortion encoding \cite{ArikanMerhav,Cohen,Merhav1999}.
%%%%%
The soft guessing under log-loss framework is similarly motivated, but with the goal of recovering a probability distribution rather than a single point estimate, making it more consistent with a fully Bayesian perspective. Allowing errors further broadens the scope of applications, enabling scenarios where the search effort can be reduced at the cost of permitting a small error probability.
%%%%%%

%%%%%%
The close connection with source coding is another key motivation of this work. Bounds on guessing moments are known to yield bounds on the normalized cumulant function of codeword lengths in corresponding variable-length source coding problems (without the prefix constraint) (see, e.g., \cite{SasonVerdu, Wu}).
%%%%%%
Naturally, the guessing framework we propose is related to a variable-length lossy source coding problem under log-loss and allowing errors.
%%%%%%
A further goal of this paper is to derive bounds on the cumulant function of codeword lengths in this setting, thereby extending prior results for the error-free case \cite{Shkel,Wu}.
%%%%%%
Next, we outline the organization of the paper and highlight our main technical contributions.
\subsection{Organization and Contributions}
%%%%%
In Section \ref{Preliminaries}, we review the definitions of the Rényi entropy, smooth Rényi entropy, and their conditional variants. We also present several key properties and derive new ones that are useful in our proofs. In particular, we extend a chain rule for smooth Rényi entropy, due to Renner and Wolf \cite{Renner05}, to the conditional case (see Lemma \ref{Lemma_chain_rule_conditional_smooth_Renyi}).

Section \ref{SectionSoftGuessingAlloingErrors} forms the core of this paper.
%%%%
Here, we first formulate the problem of soft guessing under log-loss allowing errors, and then identify the optimal guessing strategy.
%%%%%%
This strategy is based on list guessing \cite{Wu}, augmented with a probabilistic stopping rule that allows termination after a certain number of attempts with a carefully chosen probability, similar to \cite{Kuzuoka}.
We then establish single-shot upper and lower bounds on the minimal soft guessing moment in terms of the smooth Rényi entropy of a derived random variable $Z$, representing the index of the list containing the original random variable $X$ (see Theorem \ref{MainTheorem_OneShot}).
We also derive explicit bounds involving the smooth Rényi entropy of 
$X$ itself (see Proposition \ref{Proposition_OneShot}), resembling those in \eqref{eq:upper_bound_wu_joudeh} and \eqref{eq:lower_bound_wu_joudeh}. These explicit bounds are instrumental in obtaining the guessing exponent in the i.i.d. asymptotic regime.

In Section \ref{SectionGuessingSideInformation}, we extend the results of Section \ref{SectionSoftGuessingAlloingErrors} to the case where side information is available to the guesser. In the resulting bounds, the smooth Rényi entropy is replaced by a conditional smooth Rényi entropy due to Kuzuoka \cite{Kuzuoka}. 
Our new conditional chain rule in Lemma \ref{Lemma_chain_rule_conditional_smooth_Renyi} is crucial for deriving the explicit lower bound on the guessing moment.

In Section \ref{SectionConnection}, we establish a connection between the considered guessing problem and the problem of variable-length lossy source coding under log-loss and allowing errors. We show that the minimal normalized cumulant generating function of codeword lengths is bounded above and below in terms of the minimal soft guessing moment. By using this relationship and the results in Section \ref{SectionSoftGuessingAlloingErrors}, we give bounds on the minimal normalized cumulant generating function of codeword lengths in terms of the smooth R\'enyi entropy. Moreover, we investigate the tightness of these bounds.
%%%%%
\subsection{Basic Notations} \label{Notations}
%%%%%
Random variables are denoted by uppercase letters (e.g., $X$, $Y$, $Z$), while realizations of random variables are denoted by lowercase letters (e.g., $x$, $y$, $z$). The set in which a random variable takes values is denoted by the corresponding calligraphic letter, e.g., random variable $X$ takes values in the set $\mathcal{X}$. The $n$-fold Cartesian product of $\mathcal{X}$ is denoted by $\mathcal{X}^n$. Random variables take values in a finite set unless otherwise stated. We use conventional notations for probability mass functions (pmf), e.g., $P_X$, $P_{X,Y}$, and $P_{X|Y}$ denote the pmf of $X$, the joint pmf of $X, Y$, and the conditional pmf of $X$ given $Y$, respectively. The expectation operator is denoted by $\mathbb{E}[\cdot]$. The set of probability distributions on $\mathcal{X}$, i.e., probability simplex, is denoted by $\mathcal{P}(\mathcal{X})$. The cardinality of a set $\mathcal{X}$ is denoted by $|\mathcal{X}|$. For $a \in \mathbb{R}$, $\lfloor a \rfloor$ is the greatest integer less than or equal to $a$ and $\lceil a \rceil$ is the least integer greater than or equal to $a$. Throughout the paper, $\log (\cdot)$ denotes $\log_2 (\cdot)$ and $\exp(\cdot)$ denotes $2^{(\cdot)}$. 
%%%%%
\section{Preliminaries}
\label{Preliminaries}
%%%%%
The R\'enyi entropy, the smooth R\'enyi entropy, and their conditional versions play an important role in this paper. In this section, we present their definitions and some useful properties used in the proofs of the main theorems of this paper. As we see further on, in the context of guessing, R\'enyi entropy orders of most relevance are $\alpha \in (0,1)$. Hence, we assume $\alpha \in (0,1)$ throughout the paper. Moreover, we assume $\epsilon \in [0,1)$ for the smoothness parameter.
%%%%%
\subsection{R\'enyi Entropy and Arimoto-R\'enyi Conditional Entropy} \label{Definition_RE}
%%%%%
%\subsubsection{R\'enyi Entropy}
The R\'enyi entropy of order $\alpha$ is defined as \cite{Renyi} 
\begin{align}
H_{\alpha}(X) := \frac{1}{1 - \alpha} \log \sum_{x \in {\cal X}} [P_{X}(x)]^{\alpha}. \label{RenyiEntropy}
\end{align}
By using L'H\^{o}pital's rule, it can be shown that
\begin{align}
\lim_{\alpha \to 1} H_{\alpha}(X) = H(X), \label{RenyiEntropyandShannonEntropy}
\end{align}
where $H(X)$ denotes the standard Shannon entropy. 
%%%%%
For brevity, we will use the convention $H_1(X) = H(X)$, where it is understood that $H_1(X)$ is obtained by taking the limit as in \eqref{RenyiEntropyandShannonEntropy}.
%%%%%
For a pair of random variables $X$ and $Y$, the order-$\alpha$ Arimoto-R\'enyi conditional entropy of $X$ given $Y$ is defined as  \cite{Arimoto} 
\begin{align}
H_{\alpha}(X|Y) := \frac{\alpha}{1 - \alpha} \log \left ( \sum_{y \in \mathcal{Y}} \left[ \sum_{x \in \mathcal{X}} [P_{X,Y}(x,y)]^{\alpha} \right]^{1/\alpha} \right ). \label{ArimotoRenyiEntropy}
\end{align}
As in \eqref{RenyiEntropyandShannonEntropy}, we have $H_{1}(X|Y) = H(X|Y)$, which is the Shannon conditional entropy.
Moreover, it holds that 
\begin{equation}
   H_{\alpha}(X|Y) \leq H_{\alpha}(X),  
\end{equation}
with equality if $X$ and $Y$ are independent.
It should be noted that there are several distinct suggestions for formulating the conditional R\'enyi entropy which serve different purposes, see, e.g., \cite{Fehr2014}. The Arimoto-R\'enyi version is  best suited for characterizing guessing moments with side information, as shown by Ar\i kan \cite{Arikan}.
%%%%%%
\subsection{Smooth R\'enyi Entropy and Conditional Smooth R\'enyi Entropies} 
\label{Definition_SRE}
%%%%%%%
\subsubsection{Smooth R\'enyi Entropy}
%%%%%
The smooth  R\'enyi entropy was introduced by Renner and Wolf in \cite{Renner05}.
The $\epsilon$-smooth R\'enyi entropy of order $\alpha$ is defined as
\begin{align}
H^{\epsilon}_{\alpha}(X) := \frac{1}{1 - \alpha} \log \left ( \inf_{Q_X \in \mathcal{B}^{\epsilon}(P_X) } \sum_{x \in \mathcal{X}} [Q_X(x)]^{\alpha} \right ), \label{SmoothRenyiEntropy}
\end{align}
where $\mathcal{B}^{\epsilon}(P_X)$ is a set of functions $Q_X: \mathcal{X} \rightarrow [0,1]$ such that $Q_X(x) \leq P_X(x)$ for all $x \in \mathcal{X}$ and $\sum_{x \in \mathcal{X}} Q_X(x) \geq 1 -\epsilon$.
%%%%%
\begin{Remark} \label{RenyiEntropyandSmoothRenyiEntropy}
$H^{0}_{\alpha}(X)$ is equal to the R\'enyi entropy $H_{\alpha}(X)$, since $Q_X(x) = P_X(x)$ for all $x \in \mathcal{X}$ when $\epsilon=0$.
\end{Remark}
%%%%%%
For $X$ and $Y$, the joint $\epsilon$-smooth R\'enyi entropy of order $\alpha$ is defined in a similar manner as \eqref{SmoothRenyiEntropy}, and is denoted by $H^{\epsilon}_{\alpha}(X,Y)$.
%%%%
As shown in \cite{Renner05}, certain properties satisfied by the Shannon entropy have counterparts for the smooth R\'enyi entropy. We present ones that are useful to us in our proofs.  
%%%%%%%
\begin{Lemma}[{\cite[Lemma 7]{Renner05}}] 
\label{Lemma_smooth_monotonic}
    For random variables $X$ and $Y$, we have 
    \begin{align}
    H^{\epsilon}_{\alpha}(X) \leq H^{\epsilon}_{\alpha}(X, Y).
    \end{align}
\end{Lemma}
%%%%%%%%
\begin{Lemma} [{\cite[Eq.\ (12)]{Renner05}}] \label{Lemma_smooth_function}
    Let $f(X)$ be a function of $X$. Then 
    \begin{align}
    H^{\epsilon}_{\alpha}(f(X)) \leq H^{\epsilon}_{\alpha}(X). \label{InequalitySmoothRenyi}
    \end{align}
\end{Lemma}
By setting $\epsilon=0$ in \eqref{InequalitySmoothRenyi} and recalling Remark \ref{RenyiEntropyandSmoothRenyiEntropy}, we recover  $ H_{\alpha}(f(X)) \leq H_{\alpha}(X)$.
%%%%%
Moreover, setting $\alpha = 1$ 
and recalling \eqref{RenyiEntropyandShannonEntropy}, we recover $H (f(X)) \leq H (X)$, 
a well-known inequality for the Shannon entropy (see, e.g., \cite[Problem 2.4]{Cover}).
%%%%%%
\subsubsection{Smooth R\'enyi Entropy Explicit Formula}
%%%%%
Koga \cite{Koga} showed that for orders $\alpha \in (0,1)$, the infimum in the definition of smooth R\'enyi entropy can be solved explicitly. 
%%%%%
For convenience, we henceforth assume without loss of generality that
\begin{equation}
\label{eq:Px_order}
    P_X(1) \geq P_X(2) \geq \ldots \geq P_X(|\mathcal{X}|) >0.
\end{equation}
%%%%%
Given $\epsilon \in [0,1)$, let $i^*_X$ be the minimum integer in 
$\{1,2,\ldots,|\mathcal{X}| \}$ such that 
    \begin{align}
    \sum_{k=1}^{i^*_X} P_X(k) \geq 1 - \epsilon. \label{definition_i_star}
    \end{align}
    Furthermore, define $Q^\epsilon_X(j)$ such that
    \begin{align}
        Q^\epsilon_X(j)=
        \begin{cases}
            P_X(j), & j=1, 2, \ldots, i^*_X - 1, \\
            1 - \epsilon - \sum_{i=1}^{i^*_X - 1} P_X(i), & j=i^*_X, \\
            0, & j=i^*_X + 1, \ldots, |\mathcal{X}|.
        \end{cases}
    \end{align}
%%%%%%%%
\begin{Lemma}[{\cite[Theorem 1]{Koga}}]
\label{Lemma_explicit_Koga}
The smooth R\'enyi entropy is equal to 
    \begin{align}
    H^{\epsilon}_{\alpha}(X) = \frac{1}{1 - \alpha} \log \left ( \sum_{j=1}^{i^*_X} \left[Q^\epsilon_X(j) \right]^{\alpha} \right ).
    \end{align}
\end{Lemma}
%%%%%%%
\begin{Remark} \label{Remark_Q^epsilon_Z}
For any other random variable, e.g. $Z$, we define $i^*_Z$ and $Q^\epsilon_Z$ in a similar manner as in Lemma \ref{Lemma_explicit_Koga}.
\end{Remark}
%%%%%
\subsubsection{Renner-Wolf Conditional Smooth R\'enyi Entropy and Chain Rule}
%%%%%
Similar to conditioning in the standard R\'enyi entropy case, there are also multiple proposals for formulating conditional versions of the $\epsilon$-smooth R\'enyi entropy.
%%%%%
The Renner-Wolf conditional smooth R\'enyi entropy of $X$ given $Y$ is defined as \cite{Renner05} 
\begin{align}
\tilde{H}^{\epsilon}_{\alpha}(X|Y) := \frac{1}{1 - \alpha} \log \left ( \inf_{Q_{X,Y} \in \mathcal{B}^{\epsilon}(P_{X,Y}) } \max_{\substack{ y \in \mathcal{Y} :\\ P_Y(y) > 0}} \sum_{x \in \mathcal{X}} \left[\frac{Q_{X,Y}(x,y)}{P_Y(y)} \right]^{\alpha} \right ), \label{Definition_ConditionalSRE}
\end{align}
where $\mathcal{B}^{\epsilon}(P_{X,Y})$ is a set of functions $Q_{X,Y}: \mathcal{X} \times \mathcal{Y} \rightarrow [0,1]$ such that $Q_{X,Y}(x,y) \leq P_{X,Y}(x,y)$ for all $x \in \mathcal{X}$, $y \in \mathcal{Y}$, and $\sum_{x \in \mathcal{X}, y \in \mathcal{Y}} Q_{X,Y}(x,y) \geq 1 -\epsilon$.
%%%%%
A main utility of this conditional smooth R\'enyi entropy is that it enables a certain chain rule, which is particularly useful in proving the converse to our main result. 
%%%%%
\begin{Lemma} [{\cite[Lemma 5]{Renner05}}] \label{Lemma_smooth_chain_rule}
    Let $\epsilon' \geq 0$. For random variables $X$ and $Z$, 
    \begin{align}
    H^{\epsilon' + \epsilon}_{\alpha}(X, Z) \leq \tilde{H}^{\epsilon'}_{\alpha}(X|Z) + H^{\epsilon}_{\alpha}(Z).
    \end{align}
\end{Lemma}
%%%%%%
As it turns out, a special case of the above lemma, where $\epsilon' = 0$, is what we need in the present paper.
%%%%%%
\begin{Corollary}\label{Corollary_smooth_chain_rule}
For random variables $X$ and $Z$, 
\begin{align}
H^{\epsilon}_{\alpha}(X, Z) \leq \tilde{H}^{0}_{\alpha}(X|Z) + H^{\epsilon}_{\alpha}(Z),
\end{align}
where 
\begin{align}
\tilde{H}^{0}_{\alpha}(X|Z)
= \max_{\substack{ z \in \mathcal{Z} :\\ P_Z(z) > 0}} \frac{1}{1 - \alpha} \log  \sum_{x \in \mathcal{X}} \left[P_{X|Z}(x|z) \right]^{\alpha}. \label{Definition_ConditionalSRE_epsilon=0}
\end{align}
\end{Corollary}
%%%%%
Before we proceed, we note that \eqref{Definition_ConditionalSRE_epsilon=0} coincides with the third suggestion for the conditional Rényi entropy in \cite{Fehr2014}.
%%%%%
\subsubsection{Kuzuoka Conditional Smooth R\'enyi Entropy}
%%%%%
We now present a second version of the conditional smooth R\'enyi entropy proposed by Kuzuoka \cite{Kuzuoka}.
%%%%%
This version is particularly useful for studying guessing allowing errors in the presence of side information, and can be seen as the smooth counterpart to the Arimoto-Rényi conditional entropy. 
%%%%%
Kuzuoka's version of the conditional smooth R\'enyi entropy of $X$ given $Y$ is defined as \cite{Kuzuoka}
\begin{align}
H^{\epsilon}_{\alpha}(X|Y) := \frac{\alpha}{1 - \alpha} \log \left ( \inf_{Q_{X,Y} \in \mathcal{B}^{\epsilon}(P_{X,Y}) } \sum_{y \in \mathcal{Y}} \left[ \sum_{x \in \mathcal{X}} [Q_{X,Y}(x,y)]^{\alpha} \right]^{1/\alpha} \right ). \label{KuzuokaConditionalSmoothRenyiEntropy}
\end{align}
\begin{Remark} \label{ConditionalRenyiEntropyandSmoothRenyiEntropy}
    When $\epsilon=0$ in \eqref{KuzuokaConditionalSmoothRenyiEntropy}, we see that $H^{0}_{\alpha}(X|Y)$ is equal to the Arimoto-R\'enyi conditional entropy $H_{\alpha}(X|Y)$.
\end{Remark}
%%%%%%
The following lemma shows that Kuzuoka's version of the conditional smooth R\'enyi entropy satisfies monotonicity.
%%%%%%
\begin{Lemma} \label{Lemma_monotonicity_conditional_smooth_Renyi}
    For random variables $X$, $Y$, and $Z$, 
    \begin{align}
    H^{\epsilon}_{\alpha}(X|Y) \leq H^{\epsilon}_{\alpha}(X, Z | Y). \label{monotonicity_conditional_smooth_Renyi}
\end{align}
\end{Lemma}
%%%%%
The above lemma can be shown using the same argument in the proof of \cite[Proposition 4]{Fehr2014}, where the above monotonicity property is shown to hold for the Arimoto-R\'enyi conditional entropy (i.e., $\epsilon = 0$).
%%%%%%
\subsubsection{Conditional Smooth R\'enyi Entropy Explicit Formula} 
%%%%%
Building on Koga's explicit formula, Kuzuoka \cite{Kuzuoka} showed a similar explicit formulation for his conditional smooth R\'enyi entropy of order $\alpha \in (0,1)$.
%%%%%
To present this, for each $y \in \mathcal{Y}$, let $(x_y^1, x_y^2, \ldots, x_y^{|\mathcal{X}|})$ be a permutation of $\mathcal{X}$ such that 
\begin{align}
    P_{X|Y}(x_y^1 | y) \geq P_{X|Y}(x_y^2 | y) \geq \cdots \geq P_{X|Y}(x_y^{|\mathcal{X}|} | y).
\end{align}
Given $\epsilon_y \in [0, 1)$ for every $y \in \mathcal{Y}$, let $i_{X|y}^*$ be the minimum integer in $\{1,2,\ldots,|\mathcal{X}| \}$ such that 
\begin{align}
\sum_{k=1}^{i_{X|y}^*} P_{X|Y}(x_y^k | y) \geq 1 - \epsilon_y.
\end{align}
Furthermore, define $Q^{\epsilon_y}_{X|Y}(x_y^j | y)$ such that 
\begin{align}
    Q^{\epsilon_y}_{X|Y}(x_y^j | y) :=
    \begin{cases}
        P_{X|Y}(x_y^j | y), & j=1, 2, \ldots, i_{X|y}^* - 1, \\
        1 - \epsilon_y - \sum_{i=1}^{i_{X|y}^* - 1} P_{X|Y}(x_y^i | y), & j=i_{X|y}^*, \\
        0, & j=i_{X|y}^* + 1, \ldots, |\mathcal{X}|.
    \end{cases} \label{ConditionalQ}
\end{align}
%%%%%%%
\begin{Lemma} [{\cite[Theorem 1]{Kuzuoka}}] \label{Lemma_explicit_Kuzuoka}
Kuzuoka's conditional smooth R\'enyi entropy is equal to 
\begin{align}
H^{\epsilon}_{\alpha}(X|Y) = \inf_{(\epsilon_y) \in \mathcal{E}_{0}(\epsilon)} \frac{\alpha}{1 - \alpha} \log \left( \sum_{y \in \mathcal{Y}} P_Y(y) \left[ \sum_{j=1}^{i_{X|y}^*} \left[Q^{\epsilon_y}_{X|Y}(x_y^j | y) \right]^{\alpha} \right]^{1/\alpha} \right ),
\end{align}
where $\mathcal{E}_{0}(\epsilon)$ is the set of $(\epsilon_y)_{y \in \mathcal{Y}}$ satisfying $0 \leq \epsilon_y \leq 1$ for all $y \in \mathcal{Y}$ and 
\begin{align}
\sum_{y \in \mathcal{Y}} \epsilon_y P_Y(y)=\epsilon. \label{DefinitionMathcalE}
\end{align}
\end{Lemma}
%%%%%%
\subsubsection{Conditional Chain Rule}
%%%%%%
As we see further on, in our converse proof in the presence of side information, we require a conditional form of the chain rule in Corollary \ref{Corollary_smooth_chain_rule}.
%%%%%
We present this in the following lemma. 
\begin{Lemma} \label{Lemma_chain_rule_conditional_smooth_Renyi}
    For random variables $X$, $Y$, and $Z$, 
    \begin{align}
    H^{\epsilon}_{\alpha}(X, Z|Y) \leq \tilde{H}^{0}_{\alpha}(X|Z,Y) + H^{\epsilon}_{\alpha}(Z|Y). \label{chain_rule_conditional_smooth_Renyi_entropy}
    \end{align}
\end{Lemma}
%%%%%%
\begin{IEEEproof}
The proof relies on the explicit formulas of Koga and Kuzuoka. See Appendix \ref{proof_Lemma_chain_rule_conditional_smooth_Renyi}.
\end{IEEEproof}
Note that Renner and Wolf \cite{Renner05} showed that the chain rule $\tilde{H}^{\epsilon}_{\alpha}(X, Z|Y) \leq \tilde{H}^{0}_{\alpha}(X|Z,Y) + \tilde{H}^{\epsilon}_{\alpha}(Z|Y)$ holds for their version of the conditional smooth R\'enyi entropy (see Eq. (10) in \cite{Renner05}). This, however, does not imply th chain rule in Lemma \ref{Lemma_chain_rule_conditional_smooth_Renyi}, and more importantly, is not fit for our purpose of proving a converse under side information. 
%%%%%%%
\section{Soft Guessing Allowing Errors} \label{SectionSoftGuessingAlloingErrors}
%%%%
We now provide a formal description of the soft guessing allowing errors problem, which generalizes both the Kuzuoka \cite{Kuzuoka} and Wu-Joudeh frameworks \cite{Wu}.
%Recall that $\mathcal{X}=\{1, 2, \ldots, K\}$ and the probability order in \eqref{eq:Px_order}.
A soft guessing strategy with give-up probabilities is specified by the pair $(\mathcal{G}_{\mathrm{soft}}, \pi)$, where
\begin{align}
    &\mathcal{G}_{\mathrm{soft}}=(\hat{P}_1, \hat{P}_2, \ldots, \hat{P}_N), \quad \hat{P}_i \in \mathcal{P}(\mathcal{X}) \ \text{for all} \ i=1, 2, \ldots, N, \\
    &\pi =(\pi_1, \pi_2, \ldots, \pi_N), \quad 0 \leq \pi_i \leq 1 \ \text{for all} \ i=1, 2, \ldots, N,
\end{align}
%%%%%
for some integer $N$.
%%%%%
At the $j$-th step ($j=1, 2, \ldots, N$), the guesser first makes a randomized decision to either give up and declare an error with probability $\pi_j$, or proceed with the guessing procedure with probability $1- \pi_j$.
%%%%%%
In the latter case, the guesser asks the question ``is $d(x, \hat{P}_j) \leq D$?'', where $d(x, \hat{P}_j)$ is the log-loss defined in \eqref{Def_LogLoss} and $D \geq 0$ is a predetermined distortion level.
%%%%%%
Guessing continues until either an error is declared, or an answer ``yes" is returned.
%%%%%

Given a distortion level $D$, a strategy that terminates with probability $1$ is called $D$-admissible and is denoted by $(\mathcal{G}_{\mathrm{soft}}(D), \pi)$.
For such a strategy, guessing continues until (a) the answer is ``yes''  for some $j \in \{1,2,\ldots,N\}$, or (b) the guesser declares an error. In case of (a), the induced guessing function is $g(x) = j$, i.e., the smallest $j$ for which $d(x, \hat{P}_j) \leq D$ is satisfied.
%%%%%
Similar to \eqref{eq:lambda_i}, the  probability that the guesser does not give up before the $i$-th guess is defined as
\begin{align}
    \lambda_i := \prod_{j=1}^{i} (1-\pi_j)
\end{align}
for $i=1, 2, \ldots, N$. 
%for the sake of brevity we define $U := G(X)$, taking values in $ \{1, 2, \ldots, N\}$ with a pmf
%\begin{align}
%    P_U(u) = \sum_{x \in G^{-1}(u)} P_X(x),
%\end{align}
%where $G^{-1}(u) := \{x \in \mathcal{X} : G(x) = u \}$ is the preimage of $G$.
%%%%%
Define the corresponding stochastic guessing function $G: \mathcal{X} \to \{1, \ldots, N\} \cup \{0\}$ by
\begin{align}
G(x) := 
\begin{cases}
g(x), & \mathrm{with ~ probability}~ \lambda_{g(x)}, \\
0, & \mathrm{with ~ probability}~ 1-\lambda_{g(x)}.
\end{cases}  \label{StochasticSoftGuessingFunction}
\end{align}
%%%%%
%%%%%
The probability that a soft reconstruction for $X$ is correctly guessed at the $i$-th step ($i=1, 2, \ldots, N$) before giving up is equal to $\mathbb{P}[G(X)=i] = \lambda_i \left(\sum_{x \in g^{-1}(i)} P_X(x) \right)$, where $g^{-1}(i) := \{x \in \mathcal{X} : g(x) = i \}$.
%%%%%
On the other hand, the probability that the guesser gives up and declares an error before finding a soft reconstruction is 
\begin{equation}
    P_e := \mathbb{P}[G(X)=0]
    = 1 - \sum_{i=1}^{N} \lambda_i \left(\sum_{x \in g^{-1}(i)} P_X(x) \right).
    \label{ErrorProbability}
\end{equation}
Hence, the $\rho$-th soft guessing moment is given by
%%%%%%
\begin{align}
M_X \big(\rho,\mathcal{G}_{\mathrm{soft}}(D),\pi \big) 
:= \mathbb{E}\left[G(X)^{\rho} \right]
=  \sum_{i=1}^{N}  \lambda_i \left(\sum_{x \in g^{-1}(i)} P_X(x) \right) \times i^{\rho}. \label{definition_soft_guessing_moment}
\end{align}
%%%%%%
Under the error probability constraint that $P_e \leq \epsilon$, the minimal $\rho$-th soft guessing moment is defined as
\begin{align}
  M_X^{\star}(\rho,D,\epsilon) := \min_{\mathcal{G}_{\mathrm{soft}}(D),\pi: P_e \leq \epsilon} M_X \big(\rho,\mathcal{G}_{\mathrm{soft}}(D),\pi \big). 
  \label{FundamentalLimitSoftGuessing}
\end{align}
%%%%%

Our main object of interest is $M_X^{\star}(\rho,D,\epsilon)$. In particular, we wish to characterize the optimal soft guessing allowing errors strategy that achieves the minimization in \eqref{FundamentalLimitSoftGuessing}, and find single-shot upper and lower bounds for $M_X^{\star}(\rho,D,\epsilon)$.
%%%%%
It is clear that the settings of Kuzuoka \cite{Kuzuoka} and Wu-Joudeh  \cite{Wu} are special cases of the setting defined above, i.e., $ M_X^{\star}(\rho,D,\epsilon) $ reduces to $M_X^{\star}(\rho,\epsilon)$ defined in \eqref{eq:guessing_moment_errors} under  $D=0$, and to $M_X^{\star}(\rho,D) $ defined in \eqref{eq:guessing_moment_soft} under $\epsilon =0$.
%%%%%
Therefore, the bounds we seek should naturally recover those in \eqref{eq:upper_bound_Kuzuoka}--\eqref{eq:lower_bound_Kuzuoka} and \eqref{eq:upper_bound_wu_joudeh}--\eqref{eq:lower_bound_wu_joudeh} as special cases.
%%%%%
\begin{Remark} \label{remark_soft_guessing}
For any error probability constraint $\epsilon$ and distortion level $D$, an obvious $D$-admissible soft guessing strategy is obtained by setting $N = |\mathcal{X}|$, choosing $\hat{P}_{1}, \hat{P}_{2}, \ldots, \hat{P}_{|\mathcal{X}|}$ to be distinct hard reconstructions (i.e. single mass pmfs) covering all realization in $\mathcal{X}$, and setting all give up probabilities to zero. 
%%%%%
Since we are interested in optimal strategies that attain \eqref{FundamentalLimitSoftGuessing}, we may restrict our attention to strategies with $N \leq |\mathcal{X}|$ without any loss in generality.
\end{Remark}
%%%%%
\subsection{Bounds on Guessing Moments} \label{MainResults}
%%%%%
We now present the first result of our paper, in which we provide bounds 
on $ M_X^{\star}(\rho,D,\epsilon) $.
%%%%%
\begin{Theorem} \label{MainTheorem_OneShot}
    For any $\rho > 0$, $D \geq 0$, and $\epsilon \in [0,1)$, the guessing moment $ M_X^{\star}(\rho,D,\epsilon)$ is bounded above and below as
    \begin{align}
    M_X^{\star}(\rho,D,\epsilon) \leq \exp \left(  \rho H^{\epsilon}_{\frac{1}{1+\rho}}(Z)\right) 
    \label{MainUpperBound}
    \end{align}
    and 
    \begin{align}
    M_X^{\star}(\rho,D,\epsilon) \geq (1+\log |\mathcal{X}|)^{-\rho} \exp \left( \rho H^{\epsilon}_{\frac{1}{1+\rho}}(Z) \right) \label{MainLowerBound}
    \end{align}
    respectively, where $Z$ is defined as
    \begin{align}
        Z := \left \lceil \frac{X}{\lfloor \exp(D) \rfloor} \right \rceil. \label{DefinitionZ}
    \end{align}
\end{Theorem}
%%%%%
\begin{IEEEproof}
We first find the optimal strategy in Section \ref{SectionOptimalGuessing}, and then prove Theorem \ref{MainTheorem_OneShot} in Section \ref{subsection_proof_Theorem_1}.
\end{IEEEproof}
%%%%%%
The optimal soft guessing allowing errors strategy builds upon the Wu-Joudeh list guessing strategy \cite{Wu}, which in turn, leverages the close connection between soft reconstruction under log-loss and list decoding established by Shkel and Verd\'u  \cite{Shkel} in the context of lossy source coding. 
%%%%%%
In particular, the set of realizations $\mathcal{X}$ is partitioned into lists of size no greater than $\lfloor \exp(D) \rfloor$, and each list induces a soft reconstruction which is uniformly supported on it. Therefore, correctly guessing the list containing $X$ incurs a log-loss of no more than $D$.
%%%%%
The random variable $Z$ in the statement of Theorem \ref{MainTheorem_OneShot} can be interpreted as the index of the list containing $X$, or equivalently, the index of the soft reconstruction covering $X$.
%%%%%
Errors are allowed by identifying a ``cut-off'' list before which guessing never stops, and after which guessing stops with a non-zero probability carefully tuned to satisfy the error probability constraint of $\epsilon$. 
%%%%%
Details are presented in what follows.

We can see from Theorem \ref{MainTheorem_OneShot} that for $0 \leq D < 1$, and thus $\lfloor \exp(D) \rfloor =1$, we recover Kuzuoka's guessing allowing errors bounds in \eqref{eq:upper_bound_Kuzuoka} and \eqref{eq:lower_bound_Kuzuoka}. 
In this case, the predetermined distortion level is small, and satisfying it requires that each list consists of no more than a single realization, reducing the setting to the one in \cite{Kuzuoka}. 
%%%%%
On the other hand, if we set $\epsilon = 0$ while allowing $D$ to be arbitrary, we obtain the following bounds for soft guessing under log-loss 
\begin{align}
\label{eq:bounds_soft_guessing_epsilon_0}
(1+\log |\mathcal{X}|)^{-\rho} \exp \left( \rho H_{\frac{1}{1+\rho}}(Z) \right) \leq M_X^{\star}(\rho,D) \leq \exp \left(  \rho H_{\frac{1}{1+\rho}}(Z)\right).
\end{align}
%%%%%
Compared to the Wu-Joudeh lower bound in \eqref{eq:lower_bound_wu_joudeh}, the lower bound in \eqref{eq:bounds_soft_guessing_epsilon_0} is tighter, which can be deduced from Proposition \ref{Proposition_OneShot} presented further on in Section \ref{subsec:explicit_bounds}.
%%%%%
The upper bound in \eqref{eq:bounds_soft_guessing_epsilon_0}, however, is not directly comparable to the Wu-Joudeh upper bound in \eqref{eq:upper_bound_wu_joudeh}.
%%%%%%
We discuss this point in more detail in Section \ref{subsec:explicit_bounds}, where we derive an explicit bound that corresponds to \eqref{eq:upper_bound_wu_joudeh} in the allowing errors regime of $\epsilon > 0$; and demonstrate a partial relationship with the upper bound in \eqref{MainUpperBound}.
%%%%%
\subsection{Optimal Strategy} 
\label{SectionOptimalGuessing}
%%%%%
Motivated by the connection between guessing under log-loss and list guessing \cite{Wu}, 
we now present a strategy based on list guessing with randomized stopping, and then prove its optimality afterwards. 
We will denote this strategy by $(\mathcal{G}_{\mathrm{soft}}^\star, \pi^\star)$.
Recall that $P_X(1) \geq P_X(2) \geq \ldots \geq P_X(|\mathcal{X}|) > 0$. Moreover, define the following non-negative integers
\begin{align}
    K &:= \left \lceil \frac{i^*_X}{L} \right \rceil, \label{Definition_K} \\
    K' &:= \left \lceil \frac{|\mathcal{X}| - i^*_X}{L} \right \rceil, \\
    N &:= K + K',
\end{align}
%%%%%%%
where $L:= \lfloor \exp(D) \rfloor$.
%%%%%%
Let $\mathcal{L}_1, \ldots, \mathcal{L}_N$ be a collection of lists each of size not exceeding $L$, defined as 
\begin{align}
    \mathcal{L}_i &:= \{(i-1) L+1,  (i-1) L +2, \ldots, i L \}, ~ i=1, 2, \ldots, K-1, \label{List} \\
    \mathcal{L}_K &:= \{(K-1) L+1,  (K-1) L +2, \ldots, i^*_X \}, \\
    \mathcal{L}_{K+j} &:= \{(j-1) L+i^*_X+1,  (j-1) L + i^*_X +2, \ldots, j L +i^*_X \}, ~ j=1, 2, \ldots, K'-1, \\
    \mathcal{L}_{N} &:= \{(K'-1) L+i^*_X+1,  (K'-1) L +i^*_X+2, \ldots, |\mathcal{X}| \}.
\end{align}

From the above construction, it is clear that the above collection of lists form a partition of $\mathcal{X}$.
%%%%%%
Moreover, from the above collection of lists, we induce a sequence of $N$ soft reconstructions $\mathcal{G}_{\mathrm{soft}}^\star=(\hat{P}_1^\star, \hat{P}_2^\star, \ldots, \hat{P}_N^\star)$ such that
\begin{align}
    \hat{P}_i^\star(x) &:= 
        \begin{cases}
            \frac{1}{|\mathcal{L}_i|}, & \forall x \in \mathcal{L}_i, \\
            0, & \mathrm{otherwise},
        \end{cases} \label{OptimalSoftGuess}
\end{align}
for every $i=1, 2, \ldots, N$.
\begin{comment}
\begin{align}
    \hat{P}_i^\star(x) &:= 
        \begin{cases}
            \frac{1}{L}, & \forall x \in \mathcal{L}_i, \\
            0, & \mathrm{otherwise},
        \end{cases} \qquad (i=1, 2, \ldots, K-1) \label{OptimalSoftGuess_1_K-1} \\
    \hat{P}_K^\star(x) &:= 
        \begin{cases}
            \frac{1}{i^*_X - (K-1) L}, & \forall x \in \mathcal{L}_K,\\
            0, & \mathrm{otherwise},
        \end{cases} \\
    \hat{P}_{K+j}^\star(x) &:=
        \begin{cases}
            \frac{1}{L}, & \forall x \in \mathcal{L}'_j, \\
            0, & \mathrm{otherwise},
        \end{cases} \qquad (j=1, 2, \ldots, K'-1) \\
    \hat{P}_{N}^\star(x) &:= 
        \begin{cases}
            \frac{1}{|\mathcal{X}| - (K'-1) L-i^*_X}, & \forall x \in \mathcal{L}'_{K'} \\
            0, & \mathrm{otherwise}.
        \end{cases} \label{OptimalSoftGuess}
\end{align}
\end{comment}
Finally, we define sequences of stopping probabilities  $\pi^\star=(\pi_1^\star, \pi_2^\star, \ldots, \pi_N^\star)$ as 
\begin{align}
    \pi_j^\star := 
    \begin{cases}
        0, & j=1, 2, \ldots, K-1, \\
        1 - \frac{\sum_{i=(K-1) L+1}^{i^*_X} Q^\epsilon_X(i)}{\sum_{i=(K-1) L+1}^{i^*_X} P_X (i)}, & j=K, \\
        1, & j=K+1, K+2, \ldots, N, \label{OptimalGiveUpProbability}
    \end{cases}
\end{align}
from which we obtain
\begin{align}
    \lambda_i^\star := \prod_{j=1}^{i} (1-\pi_j^\star). \label{OptimalLambda}
\end{align}
%%%%
Note that $\mathcal{L}_K$ is the cut-off list we referred to earlier, at  which the guesser may decide to stop with non-zero probability.

Now let $g^\star(x)$ be the guessing function induced by $(\mathcal{G}_{\mathrm{soft}}^\star, \pi^\star)$. Using $g^\star(x)$ and $\lambda_i^\star$, we define a corresponding stochastic guessing function $G^\star$ similar to \eqref{StochasticSoftGuessingFunction}. It follows that the error probability $P_e$ of this strategy is exactly $\epsilon$, as seen from
\begin{align}
    P_e 
    &= \mathbb{P}\left[G^\star(X) = 0 \right] \\
    &= 1 - \sum_{i=1}^{N} \lambda^\star_i \left(\sum_{x \in (g^\star)^{-1}(i)} P_X(x) \right) \\
    &= 1 - \sum_{i=1}^{K-1} \left(\sum_{x \in (g^\star)^{-1}(i)} P_X(x) \right) - \lambda^\star_K \left(\sum_{x \in (g^\star)^{-1}(K)} P_X(x) \right) \label{err_prob_stepA} \\
    &= 1 - \sum_{i=1}^{i^*_X} Q^\epsilon_X(i) \label{err_prob_stepB} \\
    &= \epsilon, \label{OptimalErrorProbability}
\end{align}
where $(g^\star)^{-1}(i) := \{x \in \mathcal{X} :g^\star(x) = i \} = \mathcal{L}_i$; \eqref{err_prob_stepA} follows from \eqref{OptimalGiveUpProbability} and \eqref{OptimalLambda}; \eqref{err_prob_stepB} follows from the definition of $(\mathcal{G}_{\mathrm{soft}}^\star, \pi^\star)$ and the definition of $Q^\epsilon_X$ in Lemma \ref{Lemma_explicit_Koga}; and \eqref{OptimalErrorProbability} is due to the definition of $Q^\epsilon_X$.

%%%%%
It is also clear that the above guessing strategy is $D$-admissible, since for any $x \in \mathcal{X}$ there exists $\hat{P}_j^\star$ such that
\begin{align}
    d(x, \hat{P}_j^\star)
    =\log \frac{1}{\hat{P}_j^\star (x)} 
    \leq \log L 
    \leq D. \label{OptimalGuess_D-admissible}
\end{align}
Therefore, $\mathcal{G}_{\mathrm{soft}}^\star$ can be denoted by $\mathcal{G}_{\mathrm{soft}}^\star (D)$.
The following proposition shows the optimality of $(\mathcal{G}_{\mathrm{soft}}^\star (D), \pi^\star)$.

\begin{Proposition} \label{PropositionOptimalStrategy}
For every $\rho > 0$, the strategy $(\mathcal{G}_{\mathrm{soft}}^\star (D), \pi^\star)$ satisfies
    \begin{align}
        M^\star_X (\rho, D, \epsilon)= M_X \big(\rho,\mathcal{G}_{\mathrm{soft}}^\star(D),\pi^\star \big).
    \end{align} 
\end{Proposition}
\begin{IEEEproof}
    The proof is presented in Appendix \ref{ProofOptimalStrategy}.
\end{IEEEproof}
%%%%%%%
\subsection{Proof of Theorem \ref{MainTheorem_OneShot}} \label{subsection_proof_Theorem_1}
Equipped with the optimal strategy described above, we now proceed to prove Theorem \ref{MainTheorem_OneShot}. 
\subsubsection{Proof of Upper Bound (Achievability)}
Starting from the equality in Proposition \ref{PropositionOptimalStrategy}, we proceed as follows.
\begin{align}
    M^\star_X (\rho, D, \epsilon)
    &= M_X \big(\rho,\mathcal{G}_{\mathrm{soft}}^\star(D),\pi^\star \big) \\
    &= \sum_{i=1}^{N} \lambda_i^\star \left(\sum_{x \in (g^\star)^{-1}(i)} P_X(x) \right) i^{\rho} \\
    &= \sum_{i=1}^{K-1} \left(\sum_{x \in (g^\star)^{-1}(i)} P_X(x) \right) i^{\rho} + \lambda_K^\star \left(\sum_{x \in (g^\star)^{-1}(K)} P_X(x) \right) K^{\rho} \label{AchievabilityStep(a)} \\
    &= \sum_{i=1}^{K-1} \left( \sum_{j=(i-1) L+1}^{i L}  Q^\epsilon_X(j) \right) i^{\rho} + \left( \sum_{j=(K-1) L+1}^{i^*_X} Q^\epsilon_X(j) \right) K^{\rho} \label{AchievabilityStep(b)} \\
    &= \sum_{i=1}^{i_Z^*} Q^\epsilon_Z(i) i^\rho \label{AchievabilityStep(c)} \\
    &= \sum_{i=1}^{i_Z^*} Q^\epsilon_Z(i) \left( \sum_{k : k \leq i} 1 \right)^\rho \\
    &\leq \sum_{i=1}^{i_Z^*} Q^\epsilon_Z(i) \left ( \sum_{k : k \leq i} \left( \frac{Q^\epsilon_Z(k)}{Q^\epsilon_Z(i)} \right)^{\frac{1}{1+\rho}} \right )^\rho \label{AchievabilityStep(d)} \\
    &\leq \sum_{i=1}^{i_Z^*} Q^\epsilon_Z(i) \left ( \sum_{k=1}^{i_Z^*} \left( \frac{Q^\epsilon_Z(k)}{Q^\epsilon_Z(i)} \right)^{\frac{1}{1+\rho}} \right )^\rho \\
    &= \left (\sum_{j=1}^{i_Z^*} [Q^\epsilon_Z(j)]^{\frac{1}{1+\rho}} \right )^{1+\rho} \\
    &= \exp \left( \rho H^{\epsilon}_{\frac{1}{1+\rho}}(Z) \right) \label{AchievabilityStep(e)},
\end{align}
where \eqref{AchievabilityStep(a)} follows from \eqref{OptimalGiveUpProbability} and \eqref{OptimalLambda}; and \eqref{AchievabilityStep(b)} follows from the definitions of $\mathcal{G}_{\mathrm{soft}}^\star(D)$, $\pi^\star$, and $Q^\epsilon_X$.
%%%%%
To obtain \eqref{AchievabilityStep(c)}, we recall from Remark \ref{Remark_Q^epsilon_Z} that $i^*_Z$ and $Q^\epsilon_Z$ are defined in a similar manner to $i^*_X$ and $Q^\epsilon_X$. Since $Z$ is defined by \eqref{DefinitionZ}, it holds that
\begin{align}
    i^*_Z = \left \lceil \frac{i^*_X}{L} \right \rceil = K
\end{align}
and
\begin{align}
    Q^\epsilon_Z (i)=
    \begin{cases}
        \sum_{j=(i-1) L+1}^{i L}  Q^\epsilon_X(j), & i=1, 2, \ldots, i^*_Z - 1, \\
        \sum_{j=(i^*_Z-1) L+1}^{i^*_X} Q^\epsilon_X(j), & i=i^*_Z, \\
        0, & i=i^*_Z + 1, \ldots, \left \lceil \frac{|\mathcal{X}|}{L} \right \rceil,
    \end{cases}
\end{align}
from which \eqref{AchievabilityStep(c)} directly follows;
\eqref{AchievabilityStep(d)} follows from the definition of $Q^\epsilon_Z$; and \eqref{AchievabilityStep(e)} is due to Lemma \ref{Lemma_explicit_Koga}.
%%%%%%
\subsubsection{Proof of Lower Bound (Converse)}
For the lower bound, we use the following lemma introduced in \cite{Arikan}.
%%%%%%
\begin{Lemma} [{\cite[Lemma 1]{Arikan}}] \label{LemmaArikan}
    For non-negative numbers $a_i$ and $b_i$ ($i=1, 2, \ldots, N$) and any $\theta \in (0, 1)$, we have
    \begin{align}
        \sum_{i=1}^N a_i b_i \geq \left (\sum_{i=1}^N a_i^{\frac{-\theta}{1-\theta}} \right)^{\frac{1-\theta}{-\theta}} \left (\sum_{i=1}^N b_i^\theta \right )^{\frac{1}{\theta}}. \label{ConverseLemma}
    \end{align}
\end{Lemma}
%%%%%%
Starting again from Proposition \ref{PropositionOptimalStrategy}, and using Lemma \ref{LemmaArikan}, we obtain
\begin{align}
    M^\star_X (\rho, D, \epsilon)
    &= M_X \big(\rho,\mathcal{G}_{\mathrm{soft}}^\star(D),\pi^\star \big) \\
    &=\sum_{i=1}^N  \lambda_i^\star \left(\sum_{x \in (g^\star)^{-1}(i)} P_X(x) \right) i^\rho \\
    &\geq \left (\sum_{i=1}^N i^{-1} \right)^{-\rho} \left (\sum_{i=1}^N \left[ \lambda_i^\star \left(\sum_{x \in (g^\star)^{-1}(i)} P_X(x) \right) \right]^{\frac{1}{1+\rho}} \right )^{1+\rho} \label{converse_thorem_proof_stepA} \\
    &\geq (1+\log |\mathcal{X}|)^{-\rho} \left (\sum_{j=1}^{i_Z^*} [Q^\epsilon_Z(j)]^{\frac{1}{1+\rho}} \right )^{1+\rho} \label{converse_thorem_proof_stepB} \\
    &= (1+\log |\mathcal{X}|)^{-\rho} \exp \left( \rho H^{\epsilon}_{\frac{1}{1+\rho}}(Z) \right), \label{Converse1}
\end{align}
where \eqref{converse_thorem_proof_stepA} holds because we set $a_i = i^\rho$, $b_i = \lambda_i^\star \left(\sum_{x \in (g^\star)^{-1}(i)} P_X(x) \right)$, and $\theta=1/(1+\rho)$ in \eqref{ConverseLemma}; and \eqref{converse_thorem_proof_stepB} holds because $\sum_{i=1}^N i^{-1} \leq 1 + \log N \leq 1 + \log |\mathcal{X}|$ (which can be seen from the optimal strategy and Remark \ref{remark_soft_guessing}).
%%%%%
\subsection{Explicit Bounds}
\label{subsec:explicit_bounds}
%%%%%
The bounds on $M_X^{\star}(\rho,D,\epsilon)$  in Theorem \ref{MainTheorem_OneShot} are expressed in terms of the derived random variable $Z$. It is also of interest to find more explicit bounds that relate to the original random variable $X$. Such bounds are derived next. 
%%%%%
\begin{Proposition} \label{Proposition_OneShot}
For any $\rho > 0$, $D \geq 0$, and $\epsilon \in [0,1)$, the guessing moment 
$ M_X^{\star}(\rho,D,\epsilon)$ is bounded as
    \begin{align}
        M_X^{\star}(\rho,D,\epsilon) \leq 
       1-\epsilon + 2^\rho \exp \left ( \rho H^{\epsilon}_{\frac{1}{1+\rho}}(X) - \rho \log \lfloor \exp(D) \rfloor \right )
       \label{Achievability}
    \end{align}
    and 
    \begin{align}
        M_X^{\star}(\rho,D,\epsilon) \geq (1+\log |\mathcal{X}|)^{-\rho} \exp \left ( \rho H^{\epsilon}_{\frac{1}{1+\rho}}(X) - \rho \log \lfloor \exp(D) \rfloor \right ).
        \label{Converse}
    \end{align}
    For $0 \leq D < 1$ (i.e., $\lfloor \exp(D) \rfloor=1$), the upper bound can be strengthened to \eqref{eq:upper_bound_Kuzuoka}, while the lower bound reduces to \eqref{eq:lower_bound_Kuzuoka}.
\end{Proposition}
%%%%%%
\begin{IEEEproof}
The upper bound in \eqref{Achievability} generalizes the upper bound in \eqref{eq:upper_bound_wu_joudeh} by Wu and Joudeh \cite{Wu}, and is similarly obtained using an inequality by Bunte and Lapidoth \cite{Bunte}.
%%%%%
The lower bound in \eqref{Converse} is obtained by weakening  \eqref{MainLowerBound} using the chain rule of the Renner-Wolf
conditional smooth Rényi entropy (i.e., Corollary \ref{Corollary_smooth_chain_rule}). 
Details of the proof are presented in Appendix \ref{Appendix:Proof_proposition_explicit_bounds}.
\end{IEEEproof}
%%%%%%
Proposition \ref{Proposition_OneShot} explicitly bounds $M_X^{\star}(\rho,D,\epsilon)$ in terms of the smooth Rényi entropy of $X$ and the allowed log-loss distortion level $D$. These bounds are particularly useful for obtaining a tight asymptotic characterization of the guessing exponent, as we see further on.
%%%%%
It is clear that by setting $\epsilon=0$ in Proposition \ref{Proposition_OneShot}, and recalling Remark \ref{RenyiEntropyandSmoothRenyiEntropy}, we recover \eqref{eq:upper_bound_wu_joudeh} and \eqref{eq:lower_bound_wu_joudeh}.
%%%%%%

From the proof of Proposition \ref{Proposition_OneShot}, it is immediate that the explicit lower bound in \eqref{Converse} is looser than its counterpart in  \eqref{MainLowerBound}. 
%%%%%%
As for \eqref{Achievability}, we observe that this upper bound is looser than its counterpart in \eqref{MainUpperBound} 
in the  regime $0 \leq D < 1$ (i.e., $\lfloor \exp(D) \rfloor = 1$).
%%%%%
Beyond this regime, however, it is not immediate to see which of the two bounds is tighter. 
%%%%%
This is also not clear from the proof of  \eqref{Achievability}, which takes a slightly different route to the proof of \eqref{MainUpperBound}, making it difficult to directly compare the two.
%%%%%%
Note that even in the case $\epsilon = 0$, i.e., the Wu-Joudeh soft guessing setting, this question is still open. 

Our next result provides a partial answer to the above question in a regime beyond  $0 \leq D < 1$.
%%%%%%%
\begin{Proposition}
\label{proposition_upperComparison}
Let $\rho > 0$ and suppose that $D$ satisfies $1 < \lfloor 2^D \rfloor \leq 2$. Then
    \begin{align}
    \exp \left(  \rho H^{\epsilon}_{\frac{1}{1+\rho}}(Z)\right) \leq 1 - \epsilon + 2^\rho \exp \left(\rho H^{\epsilon}_{\frac{1}{1+\rho}} (X) - \rho \log \lfloor \exp(D) \rfloor \right). 
    \label{eq:upperComparison}
    \end{align}    
\end{Proposition}
%%%%%%
\begin{IEEEproof}
    Recall that $\log(\cdot)$ and $\exp(\cdot)$ have a common base of $2$.
    To prove \eqref{eq:upperComparison}, it suffices to prove
    \begin{align}
        H^{\epsilon}_{\frac{1}{1+\rho}}(Z) \leq \frac{1}{\rho} \log \left[1 - \epsilon + 2^\rho \exp \left(\rho H^{\epsilon}_{\frac{1}{1+\rho}} (X) - \rho \log \lfloor \exp(D) \rfloor \right) \right].
    \end{align}
    This is shown as follows:
\begin{align}
     &\frac{1}{\rho} \log \left[1 - \epsilon + 2^\rho \exp \left(\rho H^{\epsilon}_{\frac{1}{1+\rho}} (X) - \rho \log \lfloor \exp(D) \rfloor \right) \right] - H^{\epsilon}_{\frac{1}{1+\rho}}(Z) \nonumber \\
     & \quad = \frac{1}{\rho} \log \left[1 - \epsilon + \exp \left(\rho H^{\epsilon}_{\frac{1}{1+\rho}} (X) +\rho - \rho \log \lfloor \exp(D) \rfloor \right) \right] - H^{\epsilon}_{\frac{1}{1+\rho}}(Z) \\
     & \quad \geq \frac{1}{\rho} \log \left[\exp \left(\rho H^{\epsilon}_{\frac{1}{1+\rho}} (X) +\rho - \rho \log \lfloor \exp(D) \rfloor \right) \right] - H^{\epsilon}_{\frac{1}{1+\rho}}(Z) \\
     & \quad =  H^{\epsilon}_{\frac{1}{1+\rho}} (X) - H^{\epsilon}_{\frac{1}{1+\rho}} (Z) + 1 - \log \lfloor \exp(D) \rfloor \\
     & \quad \geq 0,
\end{align}
where the final inequality is due to Lemma \ref{Lemma_smooth_function} and the assumption $\lfloor 2^D \rfloor \leq 2$.
\end{IEEEproof}
%%%%%%
Beyond the above regimes, we conjecture that  \eqref{MainUpperBound} is tighter than \eqref{Achievability} in general. While we currently have no formal proof for this, numerical examples presented further on in Section \ref{subsec:source_coding_error_free} suggest that this may indeed be the case. 
%%%%%%
\subsection{Asymptotic Analysis} \label{subsection_asymptotic}
%%%%%%
In this subsection, we investigate the $\rho$-th guessing moment for i.i.d. sources. Let $X^n = (X_1, X_2, \ldots, X_n)$ be $n$ independent copies of $X$. As in \cite{Shkel} and \cite{Wu}, for $n$-letter setting, the log-loss is defined as
\begin{align}
\label{eq:n_letter_log_loss}
    d_n (x^n, \hat{P}_n) := \frac{1}{n} \log \frac{1}{\hat{P}_n (x^n)}, 
\end{align}
where $\hat{P}_n \in \mathcal{P}(\mathcal{X}^n)$. Note that this reduces to \eqref{Def_LogLoss} in the single-shot case (i.e. $n = 1$).
%%%%%%
Before we state our asymptotic bounds on the guessing moment, we first review a previous result by Sakai and Tan \cite{Sakai}. Let $H(X)$, $V(X)$, and $T(X)$ be 
    \begin{align}
    H(X) &:= \mathbb{E} \left [ \log \frac{1}{P_X (X)} \right], \label{ShannonEntropy} \\
    V(X) &:= \mathbb{E} \left [ \left( \log \frac{1}{P_X (X)} - H(X) \right)^2 \right], \\
    T(X) &:= \mathbb{E} \left [ \left | \log \frac{1}{P_X (X)} - H(X) \right |^3 \right].
    \end{align}
In \cite{Sakai}, an asymptotic expansion of the smooth R\'enyi entropy $H^{\epsilon}_{\alpha}(X^n)$ up to the third-order term was derived. 
%%%%%%
\begin{Lemma} [{\cite[Theorem 1]{Sakai}}] \label{Lemma_Asymptotic}
    Fix $\alpha \in (0,1)$ and $\epsilon \in (0,1)$. If  $V(X) = 0$, then
    \begin{align}
    H^{\epsilon}_{\alpha}(X^n) = n H(X) + O(1).
    \end{align}
    Otherwise, if $0 < V(X) < \infty$ and $T(X) < \infty$, then we have 
    \begin{align}
    H^{\epsilon}_{\alpha}(X^n) = n H(X) - \sqrt{n V(X)} \Phi^{-1}(\epsilon) - \frac{1}{2(1-\alpha)} \log n + O(1),
    \end{align}
    where $\Phi^{-1}: (0,1) \to \mathbb{R}$ is the inverse of the Gaussian cumulative distribution function $\Phi(u) = \int_{-\infty}^{u} \frac{1}{\sqrt{2 \pi}} e^{-\frac{t^2}{2}} \mathrm{d}t.$
\end{Lemma}
%%%%%%
Next, we present an asymptotic expansion for $\frac{1}{n} \log M_{X^n}^{\star}(\rho,D,\epsilon)$. In our asymptotic results, we assume that $D$ satisfies $0 \leq D < H(X)$. A justification for this assumption is provided in Remark \ref{remark_D_max} below. 
\begin{Proposition}
\label{MainAsymptotic}
Let $X^n = (X_1, X_2, \ldots, X_n)$ be $n$ independent copies of $X$, and suppose that $V(X) < \infty$ and $T(X) < \infty$. For any $\rho > 0$, $\epsilon \in (0,1)$, and $0 \leq D < H(X)$, it holds that
\begin{align}
   \frac{1}{n}\log M_{X^n}^{\star}(\rho,D,\epsilon) =  \rho \left(H(X) - D - \sqrt{\frac{V(X)}{n}} \Phi^{-1}(\epsilon) + O\left(\frac{\log n}{n} \right) \right). \label{asymptotic_result_2}
\end{align}
\end{Proposition}
%%%%%%
\begin{IEEEproof}
From Proposition \ref{Proposition_OneShot}, we obtain the following $n$-letter bounds on the moment
\begin{align}
   M_{X^n}^{\star}(\rho,D,\epsilon) \leq 1-\epsilon + 2^\rho \exp \left ( \rho H^{\epsilon}_{\frac{1}{1+\rho}}(X^n) - \rho \log \lfloor \exp(nD) \rfloor \right )
\end{align}
and 
\begin{align}
    M_{X^n}^{\star}(\rho,D,\epsilon) \geq (1+\log |\mathcal{X}^n|)^{-\rho} \exp \left ( \rho H^{\epsilon}_{\frac{1}{1+\rho}}(X^n) - \rho \log \lfloor \exp(nD) \rfloor \right ).
\end{align}
By using Lemma \ref{Lemma_Asymptotic} and the inequality $nD-1 \leq \log \lfloor \exp(nD) \rfloor \leq nD$, we get \eqref{asymptotic_result_2}.
\end{IEEEproof}
%%%%%%%
From Proposition \ref{MainAsymptotic}, it directly follows that the $\rho$-th guessing exponent for the setting we study is given by
\begin{align}
 \lim_{n \to \infty}  \frac{1}{n}\log M_{X^n}^{\star}(\rho,D,\epsilon) =  \rho \left(H(X) - D \right).
\end{align}
Note that $H(X) - D$ is the rate distortion function under the log-loss (see, e.g., \cite[Example 2]{CourtadeWesel} and \cite[Eq.\ (8)]{Shkel}).
%%%%%
Next, we provide a justification for the assumption $0 \leq D < H(X)$ in Proposition \ref{MainAsymptotic}.
%%%%%
\begin{Remark} \label{remark_D_max}
For $D \geq H(X)$, we get  $\frac{1}{n} \log M_{X^n}^{\star}(\rho,D,\epsilon) \to 0$. In this case, a single soft reconstruction, or list of size $L \to \exp(n H(X))$, is asymptotically sufficient for achieving $D$ and $\epsilon$.
%%%%%
Another justification comes from asymptotic analyses of lossy source coding, where similar assumptions are commonly made.
%%%%%
For instance, in the asymptotic analysis of variable-length lossy source coding allowing errors\cite[Section III-E]{Kostina}, it is assumed that the distortion level $D$ satisfies 
    \begin{align}
        D < D_{\mathrm{max}} := \inf_{\hat{s}} \mathbb{E}[d(S,\hat{s})], \label{def_D_max}
    \end{align}
    where $d : \mathcal{S} \times \hat{\mathcal{S}} \to [0, +\infty]$ is a distortion measure between a source symbol $s \in \mathcal{S}$ and a reproduction symbol $\hat{s} \in \hat{\mathcal{S}}$, and the expectation in \eqref{def_D_max} is with respect to the unconditional distribution of $S$. If we calculate \eqref{def_D_max} for the setup we study, we get
    \begin{align}
        D_{\mathrm{max}} 
        &= \inf_{\hat{P}} \mathbb{E}[d(X,\hat{P})] \\
        &= \inf_{\hat{P}} \mathbb{E} \left[ \log \frac{1}{\hat{P}(X)} \right] \label{calc_D_max_stepA} \\
        & =  H(P_X) + \inf_{\hat{P}}  D_{\mathrm{KL}}(P_X \| \hat{P}) \\
        %&=\mathbb{E} \left[ \log \frac{1}{P_X(X)} \right] \label{calc_D_max_stepB} \\
        &=H(P_X),
    \end{align}
    where $H(P_X) = H(X)$, and $D_{\mathrm{KL}}(P_X \| \hat{P})$ denotes Kullback–Leibler (KL) divergence between $P_X$ and $\hat{P}$. Therefore, the condition $D < H(X)$ in Proposition \ref{MainAsymptotic} corresponds to the condition $D < D_{\mathrm{max}}$.
    %where \eqref{calc_D_max_stepA} holds because we assume the log-loss \eqref{Def_LogLoss}; and \eqref{calc_D_max_stepB} is due to the non-negativity of KL divergence $\mathrm{KL}(P_X \| \hat{P}) \geq 0$, where $\mathrm{KL}(P_X \| \hat{P})$ denotes the  KL divergence between $P_X$ and $\hat{P}$. Therefore, the condition $D < H(X)$ in Proposition \ref{MainAsymptotic} corresponds to the condition $D < D_{\mathrm{max}}$.
\end{Remark}
%%%%%%%
\section{Soft Guessing Allowing Errors with Side Information} \label{SectionGuessingSideInformation}
\label{section_Side_Information}
%%%%%%
In this section, we extend the results of the previous section to the case where side information is available to the guesser. 
%%%%%%
Let $Y$ denote the side information random variable, which takes values in a finite set $\mathcal{Y}$. A soft guessing strategy with side information and give-up probabilities is specified by $\left(\mathcal{G}_{\mathrm{soft}}^y, \pi^y \right)_{y \in \mathcal{Y}}$, where for each $y \in \mathcal{Y}$ we have 
\begin{align}
    &\mathcal{G}_{\mathrm{soft}}^y=(\hat{P}_1^y, \hat{P}_2^y, \ldots, \hat{P}_{N_y}^y), \quad \hat{P}_i^y \in \mathcal{P}
    (\mathcal{X}) \ \text{for all} \ i=1, 2, \ldots, N_y, \\
    &\pi^y=(\pi_1^y, \pi_2^y, \ldots, \pi_{N_y}^y), \quad 0 \leq \pi_i^y \leq 1 \ \text{for all} \ i=1, 2, \ldots, N_y,
\end{align}
for some integer $N_y$. A $D$-admissible guessing strategy with side information is denoted by $\left(\mathcal{G}_{\mathrm{soft}}^y(D), \pi^y \right)_{y \in \mathcal{Y}}$.
%%%%%%

For any given realization of the side information $Y=y$, the guesser seeks to find a soft reconstruction for $X$ using the strategy $(\mathcal{G}_{\mathrm{soft}}^y(D), \pi^y)$ in a similar manner to the soft guessing allowing errors setting studied in Section \ref{SectionSoftGuessingAlloingErrors}. For $y \in \mathcal{Y}$, we denote the induced guessing function by $g_y(x)$ and define the probability
\begin{align}
    \lambda_i^y := \prod_{j=1}^{i} (1-\pi_j^y)
\end{align}
for $i=1, 2, \ldots, N_y$. Similar to \eqref{StochasticSoftGuessingFunction}, define a stochastic mapping $G: \mathcal{X} \times \mathcal{Y} \to \{1, \ldots, N_y \} \cup \{0\}$ by
\begin{align}
G(x|y) := 
\begin{cases}
g_y(x), & \mathrm{with ~ probability}~ \lambda^y_{g_y(x)}, \\
0, & \mathrm{with ~ probability}~ 1-\lambda^y_{g_y(x)}.
\end{cases}  
\end{align}
Given a $D$-admissible guessing strategy $\left(\mathcal{G}_{\mathrm{soft}}^y(D), \pi^y \right)_{y \in \mathcal{Y}}$, the average error probability $P_e^{\mathrm{avg}}$ (i.e. the error probability averaged over the side information $Y$) is defined as
\begin{align}
    P_e^{\mathrm{avg}}
    &= \mathbb{P}[G(X|Y)=0] \\
    &= \sum_{y \in \mathcal{Y}} P_Y(y) \left \{ 1 - \sum_{i=1}^{N_y} \lambda_i^y \left( \sum_{x \in g^{-1}_y (i)} P_{X|Y}(x|y) \right) \right \} \\
    &= 1 - \sum_{y \in \mathcal{Y}} P_Y(y) \sum_{i=1}^{N_y} \lambda_i^y \left( \sum_{x \in g^{-1}_y (i)} P_{X|Y}(x|y) \right),
\end{align}
where $g^{-1}_y(i) := \{x \in \mathcal{X} : g_y(x) = i \}$. The $\rho$-th soft guessing moment with side information is defined as
\begin{align}
    M_{X|Y} \left(\rho,\left(\mathcal{G}_{\mathrm{soft}}^y(D), \pi^y \right)_{y \in \mathcal{Y}} \right) 
    := \mathbb{E}[G(X|Y)^\rho]
    = \sum_{y \in \mathcal{Y}} P_Y(y)\sum_{i=1}^{N_y} \lambda_i^y \left( \sum_{x \in g^{-1}_y (i)} P_{X|Y}(x|y) \right) \times i^\rho.
\end{align}
Under the constraint $P_e^{\mathrm{avg}} \leq \epsilon$, the minimal $\rho$-th soft guessing moment with side information is defined as
\begin{align}
    M_{X|Y}^{\star}(\rho,D,\epsilon) := \min_{\left(\mathcal{G}_{\mathrm{soft}}^y(D), \pi^y \right)_{y \in \mathcal{Y}}: P_e^{\mathrm{avg}} \leq \epsilon}M_{X|Y} \left(\rho,\left(\mathcal{G}_{\mathrm{soft}}^y(D), \pi^y \right)_{y \in \mathcal{Y}} \right). \label{fundamental_limit_soft_guessing_side_information}
\end{align}
%%%%%
\subsection{Bounds on Guessing Moments with Side Information} 
%%%%%
We now derive bounds on $M_{X|Y}^{\star}(\rho,D,\epsilon)$, extending the bounds without side information derived in Theorem \ref{MainTheorem_OneShot}.
%%%%%
\begin{Theorem} \label{MainTheorem_OneShot_side}
    For any $\rho > 0$, $D \geq 0$, and $\epsilon \in [0,1)$, the guessing moment $M_{X|Y}^{\star}(\rho,D,\epsilon)$ is bounded above and below as
    \begin{align}
    M_{X|Y}^{\star}(\rho,D,\epsilon) \leq \exp \left(  \rho H^{\epsilon}_{\frac{1}{1+\rho}}(Z|Y)\right) 
    \label{MainUpperBound_side}
    \end{align}
    and 
    \begin{align}
    M_{X|Y}^{\star}(\rho,D,\epsilon) \geq (1+\log |\mathcal{X}|)^{-\rho} \exp \left( \rho H^{\epsilon}_{\frac{1}{1+\rho}}(Z|Y) \right). \label{MainLowerBound_side}
    \end{align}
\end{Theorem}

\begin{IEEEproof}
Fix $\rho$, $D$ and $\epsilon$. The proof relies on the following observation, which we shall prove shortly:
\begin{align}
    M_{X|Y}^{\star}(\rho,D,\epsilon) = \inf_{(\epsilon_y)\in \mathcal{E}_0(\epsilon )}\sum_{y \in \mathcal{Y}}P_{Y}(y) M_{X|Y = y}^{\star} (\rho, D, \epsilon_y), \label{relation_optimal_moments_side_information}
\end{align}
where $\mathcal{E}_0(\epsilon )$ is as defined in Lemma \ref{Lemma_explicit_Kuzuoka}, while  $M^\star_{X|Y=y} (\rho, D, \epsilon_y)$ is the minimal guessing moment given $Y = y$, defined for every $y $ and $\epsilon_y $ as in \eqref{FundamentalLimitSoftGuessing}. 
%%%%%%%
From Theorem \ref{MainTheorem_OneShot}, and the explicit form of the smooth Rénye entropy in Lemma \ref{Lemma_explicit_Koga}, it follows that $M_{X|Y = y}^{\star} (\rho, D, \epsilon_y)$ on the right-hand side of \eqref{relation_optimal_moments_side_information} is upper and lower bounded  as 
\begin{equation}
(1+\log |\mathcal{X}|)^{-\rho} \left (\sum_{j=1}^{i_{Z|y}^*} [Q^{\epsilon_y}_{Z|Y}(j|y)]^{\frac{1}{1+\rho}} \right )^{1+\rho}  \leq M_{X|Y = y}^{\star} (\rho, D, \epsilon_y) \leq \left (\sum_{j=1}^{i_{Z|y}^*} [Q^{\epsilon_y}_{Z|Y}(j|y)]^{\frac{1}{1+\rho}} \right )^{1+\rho}.
\end{equation}
By taking the expectation of these bounds with respect to $Y$ and then the infimum over $(\epsilon_y)\in \mathcal{E}_0(\epsilon )$, and using the explicit form of Kuzuoka's conditional smooth R\'enyi entropy in Lemma \ref{Lemma_explicit_Kuzuoka}, we obtain the bounds in Theorem \ref{MainTheorem_OneShot_side}.
%%%%%

To complete the proof, it remains to prove the statement in \eqref{relation_optimal_moments_side_information}. To this end, let 
\begin{align}
        M_{X|Y}^{\star}(\rho,D,\epsilon) = M_{X|Y} \left(\rho,\left(\mathcal{G}_{\mathrm{soft}}^{y \star}(D), \pi^{y \star} \right)_{y \in \mathcal{Y}} \right),
\end{align}
where $\left(\mathcal{G}_{\mathrm{soft}}^{y \star}(D), \pi^{y \star} \right)_{y \in \mathcal{Y}}$ denotes an optimal strategy that achieves the minimum in \eqref{fundamental_limit_soft_guessing_side_information}. Let $G^{\star}$ be the induced stochastic guessing function and $\epsilon_y^{\star} = \mathbb{P}[G^{\star}(X|y) = 0 | Y = y]$, i.e., the error probability of the optimal strategy given $Y = y$. Note that 
\begin{align}
    \sum_{y \in \mathcal{Y}} P_Y(y) \epsilon_y^{\star} = \epsilon, \label{error_prob_optimal_strategy_with_side_information}
\end{align}
as otherwise, we can find a strategy that has lower guessing moments (see \eqref{appendix_probability_upper_bound_optimal_strategy_1}--\eqref{appendix_probability_upper_bound_optimal_strategy_3}). Therefore, it holds that
\begin{align}
        M_{X|Y}^{\star}(\rho,D,\epsilon) 
        & = M_{X|Y} \left(\rho,\left(\mathcal{G}_{\mathrm{soft}}^{y \star}(D), \pi^{y \star} \right)_{y \in \mathcal{Y}} \right) \\
        & = \sum_{y \in \mathcal{Y}}P_{Y}(y) M_{X|Y=y} \left(\rho, \mathcal{G}_{\mathrm{soft}}^{y \star}(D), \pi^{y \star} \right) \\
        & \geq \sum_{y \in \mathcal{Y}}P_{Y}(y) M_{X|Y = y}^{\star} (\rho, D, \epsilon_y^{\star}) \\
        & \geq  \inf_{(\epsilon_y)\in \mathcal{E}_0(\epsilon )}\sum_{y \in \mathcal{Y}}P_{Y}(y) M_{X|Y = y}^{\star} (\rho, D, \epsilon_y).
        \label{eq:side_inf_lower_2}
\end{align}
The above inequalities can be made to hold with equality by choosing 
$(\epsilon_y^{\star})$ to achieve the infimum in \eqref{eq:side_inf_lower_2}, and 
$\left(\mathcal{G}_{\mathrm{soft}}^{y \star}(D), \pi^{y \star} \right)$ to attain $M_{X|Y = y}^{\star} (\rho, D, \epsilon_y^{\star})$ for every $y$.
%%%%%
Since these choices yield a feasible strategy, then equality must hold.
\end{IEEEproof}
%%%%%%
\subsection{Explicit Bounds}
%%%%%
In the next result, we derive explicit bounds mirroring the ones derived in Proposition \ref{Proposition_OneShot}.
%%%%%
These explicit bounds are expressed in terms of $X$ instead of $Z$, and are particularly useful for deriving guessing exponents in the i.i.d. asymptotic regime. 
%%%%%%
\begin{Proposition} \label{OneShotUpperBoundSideInformation}
    For any $\rho > 0$, $ D \geq 0$, and $\epsilon \in [0,1)$, 
    \begin{align}
        M_{X|Y}^{\star}(\rho,D,\epsilon) \leq 
        1 - \epsilon + 2^\rho \exp \left ( \rho H^{\epsilon}_{\frac{1}{1+\rho}}(X | Y) - \rho \log \lfloor \exp(D) \rfloor \right ) \label{explicit_upper_side_information}
    \end{align}
    and
    \begin{align}
        M_{X|Y}^{\star}(\rho,D,\epsilon) \geq (1+\log |\mathcal{X}|)^{-\rho} \exp \left( \rho H^{\epsilon}_{\frac{1}{1+\rho}}(X|Y) - \rho \log \lfloor \exp(D) \rfloor \right). \label{explicit_lower_side_information}
    \end{align}
    For $ 0 \leq D < 1$ (i.e., $\lfloor \exp(D) \rfloor = 1$), the upper bound can be strengthened to $M_{X|Y}^{\star}(\rho,D,\epsilon) \leq  \exp \left ( \rho H^{\epsilon}_{\frac{1}{1+\rho}}(X | Y)\right )$.
\end{Proposition}
%%%%%%
\begin{IEEEproof}
The proof follows that of Proposition \ref{Proposition_OneShot}, with the lower bound relying on the conditional chain rule for the smooth Rényi entropy (Lemma \ref{Lemma_chain_rule_conditional_smooth_Renyi}). Further details are provided in  Appendix \ref{Appendix:Proof_proposition_explicit_bounds_side_inf}.
\end{IEEEproof}
%%%%%
By setting $D=0$ in Proposition \ref{OneShotUpperBoundSideInformation}, we recover the upper and lower bounds of Kuzuoka \cite{Kuzuoka} (see \cite[Theorem 3]{Kuzuoka} and \cite[Theorem 4]{Kuzuoka}). Further, setting $\epsilon=0$ in Proposition \ref{OneShotUpperBoundSideInformation} and recalling Remark \ref{ConditionalRenyiEntropyandSmoothRenyiEntropy}, we recover the upper and lower bounds of Wu and Joudeh in \cite{Wu} (see the end of Section I-B in \cite{Wu}).
%%%%%%

%%%%%%
\subsection{Asymptotic Analysis} \label{Asymptotic_side_information}
%%%%%
We now derive an asymptotic expansion for $M_{X^n|Y^n}^{\star}(\rho,D,\epsilon)$, where $(X^n, Y^n)$ comprises $n$ independent copies of $(X, Y)$. We use an asymptotic expansion of Kuzuoka's conditional smooth R\'enyi entropy from \cite{Sakai}. Let $H(X|Y)$ and $U(X|Y)$ be 
    \begin{align}
        H(X|Y) & := \mathbb{E} \left[ \log \frac{1}{P_{X|Y}(X|Y)} \right], \\
        U(X|Y) & := \mathbb{E} \left[ \left( \log \frac{1}{P_{X|Y}(X|Y)} - H(X|Y) \right)^2 \right].
    \end{align}
%%%%%
\begin{Lemma} [{\cite[Theorem 2]{Sakai}}] \label{Lemma_Asymptotic_SideInformation}
 Fix $\alpha \in (0,1)$ and $\epsilon \in (0,1)$. If $U(X|Y)=0$, then 
    \begin{align}
    H^{\epsilon}_{\alpha}(X^n | Y^n) = n H(X|Y) + O(1).
    \end{align}
    Otherwise, if $0 < U(X|Y) < \infty$, then
    \begin{align}
    H^{\epsilon}_{\alpha}(X^n | Y^n) = n H(X|Y) + O(\sqrt{n}).
    \end{align}
\end{Lemma}
%%%%%
Combining Proposition \ref{OneShotUpperBoundSideInformation} and Lemma \ref{Lemma_Asymptotic_SideInformation}, we immediately obtain the following asymptotic result.
%%%%%
\begin{Proposition} \label{Asymptotic_SideInfo}
    Let $(X^n, Y^n)$ be $n$ independent copies of $(X, Y)$, and suppose that $U(X|Y) < \infty$.
    For any $\rho > 0$, $\epsilon \in (0,1)$, and $0 \leq D < H(X|Y)$, it holds that
    \begin{align}
\frac{1}{n} \log M_{X^n|Y^n}^{\star}(\rho,D,\epsilon) = \rho \left (H(X|Y) - D \right ) + O \left( \frac{1}{\sqrt{n}} \right).
    \end{align}
\end{Proposition}
%%%%%%%
It is obvious that from Proposition \ref{Asymptotic_SideInfo}, we characterize the corresponding $\rho$-th guessing exponent as 
\begin{align}
 \lim_{n \to \infty}  \frac{1}{n}\log M_{X^n|Y_n}^{\star}(\rho,D,\epsilon) =  \rho \left(H(X|Y) - D \right),
\end{align}
where  $H(X|Y) - D$ is the rate distortion function under the log-loss in the presence of side information \cite[Example 1]{CourtadeWesel}.
%%%%%%%
\section{Connection with Variable-Length Lossy Source Coding} \label{SectionConnection}
%%%%%%%
In this section, we establish a connection between soft guessing under log-loss allowing errors and variable-length lossy source coding.
%%%%%%
We start with a review of the key literature. 
%%%%%%
The log-loss distortion measure was introduced in the context of lossy source coding by Courtade and Wesel \cite{CourtadeWesel} (and later by Courtade and Weissman \cite{CourtadeWeissman}), who focused on multi-terminal settings in the asymptotic block-length regime.
%%%%%
Later on, Shkel and Verd\'u \cite{Shkel} established single-shot bounds for lossy source coding under log-loss. 
The study of Shkel and Verd\'u \cite{Shkel} is comprehensive as it treats fixed-length settings under both expected and excess distortion, and variable-length settings under both expected and excess codeword length. A key insight emerging from \cite{Shkel} is a connection to list decoding, which we leveraged earlier on in the context of soft guessing under log-loss.
%%%%%%

The intimate connection between guessing and source coding has been noticed in several works, e.g., \cite{ArikanMerhav, Sundaresan, Hanawal}. As noted in \cite{SasonVerdu}, the source coding setting that is most closely related to guessing is variable-length source coding \emph{without a prefix} constraint under a generalized notion of expected codeword length due to Campbell \cite{Campbell}. Note that Campbell's notion of generalized length coincides with the normalized cumulant generating function of codeword lengths \cite{CourtadeVerdu1}.
%%%%%%
More recently, Wu and Joudeh \cite{Wu} built upon this relationship and established a connection between soft guessing under log-loss and variable-length source coding, and in doing so, they extended some of the single-shot bounds in \cite{Shkel} to Campbell's generalized length setting. 
%%%%%%
In this section, we further extend this connection to the case where errors are allowed.
%%%%%%
\subsection{Variable-Length Source Coding Under Log-Loss Allowing Errors} \label{SectionVariable-lengthLossySourceCoding}
%%%%%%
In the considered source coding setting, let $X$ be a source random variable drawn from the alphabet $\mathcal{X}=\{1, 2, \ldots, |\mathcal{X}| \}$ according to the pmf $P_X$. Without loss of generality, we assume that $P_X$ satisfies the order in \eqref{eq:Px_order}.
%%%%%
Moreover, we consider a \emph{soft} reproduction alphabet given by the probability simplex $\mathcal{P}(\mathcal{X})$. A {\it variable-length lossy source code with soft reconstruction} is a pair of mappings $(f,\varphi)$ defined as follows. The encoder $f : {\cal X} \rightarrow \{ 0,1 \}^*$ is an injective mapping, where
\begin{equation}
    \{ 0,1 \}^* := \{\varnothing, 0, 1, 00, 01, 10, 11, 000, \ldots \}
\end{equation}
denotes the set of finite-length binary strings including an empty string $\varnothing$. 
%%%%%
On the other hand, the decoder is defined as $\varphi :  f(\mathcal{X})  \rightarrow \mathcal{P}(\mathcal{X})$, where $f(\mathcal{X}) := \{f(x) : x \in \mathcal{X} \}$.
%%%%%%
Note that in the considered variable-length source coding setting, \emph{no prefix} constraint is imposed on codewords.
%%%%%%%
The motivation for considering such a setting is discussed in \cite[Section I]{Kontoyiannis}, and several other works consider variants of this source coding setting, see, e.g., \cite{CourtadeVerdu1, CourtadeVerdu2, Kostina, SaitoMatsushima23, SasonVerdu, Shkel, Wu}.

For every $x \in {\cal X}$, the length of the codeword $f(x)$ is denoted by $\ell(f(x))$.
%%%%%%
Given a fixed parameter $\rho > 0$, the normalized cumulant generating function of codeword lengths (i.e., Campbell's expected generalized length) is defined as 
\begin{align}
\frac{1}{\rho} \log \mathbb{E} [\exp ( \rho \ell(f(X)) )].
\label{eq:generalized_expected_length}
\end{align}
%%%%%%
\begin{Remark} 
By using L'H\^{o}pital's rule, we obtain
\begin{align}
\lim_{\rho \to 0} \frac{1}{\rho} \log \mathbb{E} [\exp ( \rho \ell(f(X)) )] & = \mathbb{E}[\ell(f(X))],  \label{CGFandAverageCodeLength} \\
\lim_{\rho \to \infty} \frac{1}{\rho} \log \mathbb{E} [\exp ( \rho \ell(f(X)) )] & = \max_{x \in {\cal X}} \ell(f(x)).
\end{align}
\end{Remark}
%%%%%
Note that $\rho$ can be thought of as a tunable parameter through which \eqref{eq:generalized_expected_length} can be specialized to the expected codeword length, the maximum codeword length, or anything in between. The right choice of $\rho$ depends on the specific application (see, e.g., \cite[Remark 1]{SaitoMatsushima23} for further discussion on using Campbell's expected generalized length).
%%%%%

In the considered source coding scheme, we assume that errors are allowed in the form of excess distortion events in which 
the log-loss between $X$ and its reconstruction $\varphi(f (X))$ is greater than a predetermined distortion level $D$. The probability of excess distortion is defined as $\mathbb{P}[d(X, \varphi (f (X))) > D]$.
%%%%%%%
The code $(f,\varphi)$ is called a $(\Lambda, \rho, D, \epsilon)$-code if
\begin{align}
\frac{1}{\rho} \log \mathbb{E} [\exp ( \rho \ell(f(X)) )] \leq \Lambda 
\ \ \text{and} \ \ 
\mathbb{P}[d(X, \varphi (f (X))) > D] \leq \epsilon.
\end{align}
%%%%%
That is, the normalized cumulant generating function of codeword lengths is at most $\Lambda$ and the excess distortion probability is at most $\epsilon$.
%%%%%%
Given $\rho > 0$, $D \geq 0$, and $\epsilon \in [0,1)$, the object of interest is minimal expected generalized length
\begin{align}
\Lambda_{X}^{\star} (\rho, D, \epsilon) := \inf \{ \Lambda : \mbox{$\exists$ $(\Lambda, \rho, D, \epsilon)${\rm-code}} \}. \label{FundamentalLimitLossySourceCode}
\end{align}
%%%%%
It is readily seen that by setting any subset of the parameters $\rho,D,\epsilon$ to zero, we recover special cases of the above source coding setting studied previously in the literature, e.g., \cite{Kontoyiannis, CourtadeVerdu1, Shkel, Wu}.
%%%%%%
\begin{Remark}
  The setting we consider in this section can be seen as a special case of the general setting considered in \cite{SaitoMatsushima23}, under a specific distortion measure---the log-loss. In \cite[Theorem 2]{SaitoMatsushima23}, Saito and Matsushima derived upper and lower bounds on $\Lambda_{X}^{\star} (\rho, D, \epsilon)$ for a general distortion measure, but the finiteness of the source and reproduction alphabets is required in their theorem. Note that the reproduction alphabet under log-loss is the probability simplex $\mathcal{P}(\mathcal{X})$, which is not a finite set. Moreover, the achievability proof in \cite[Theorem 2]{SaitoMatsushima23} is based on stochastic encoding, and the optimal code is not explicit. In contrast, under the log-loss distortion measure, explicit optimal code construction is possible as shown by Shkel and Verd\'u  \cite{Shkel}. In the proof of Theorem \ref{TheoremConnection} presented further on, we make use of the optimal code construction in \cite{Shkel}.
\end{Remark}
%%%%%%
\subsection{Bounds on the Normalized Cumulant Generating Function}
%%%%%%
We now present the main result of this section, showing that the normalized cumulant generating function of codeword lengths $\Lambda_{X}^{\star} (\rho, D, \epsilon)$ is bounded above and below in terms of the guessing moment $M^\star_X (\rho, D, \epsilon)$.
%%%%%
\begin{Theorem} \label{TheoremConnection}
Given $\rho > 0$, $D \geq 0$, and $\epsilon \in [0,1)$, it holds that
\begin{align}
 \frac{1}{\rho} \log \left( 2^{-\rho}  M_{X}^{\star} (\rho, D, \epsilon) + \epsilon \right)
< \Lambda_{X}^{\star} (\rho, D, \epsilon) 
\leq \frac{1}{\rho} \log \left( M_{X}^{\star} (\rho, D, \epsilon) + \epsilon \right). \label{ConnectionTwoFundamentalLimits}
\end{align}
\end{Theorem}
\begin{IEEEproof}
    Theorem \ref{TheoremConnection} is proved by combining the three following observations:
    \begin{align}
        \Lambda_{X}^{\star} (\rho, D, \epsilon) &= \Lambda_{Z}^{\star} (\rho, 0, \epsilon), \label{Relation_LambdaX_LambdaZ} \\
        M^\star_X (\rho, D, \epsilon) &= M_{Z}^{\star} (\rho, 0, \epsilon), \label{Relation_CX_CZ} \\
        \frac{1}{\rho} \log \left( 2^{-\rho}  M_{Z}^{\star} (\rho, 0, \epsilon) + \epsilon \right) < \Lambda_{Z}^{\star} (\rho, 0, \epsilon) & \leq \frac{1}{\rho} \log \left( M_{Z}^{\star} (\rho, 0, \epsilon) + \epsilon \right), \label{Relation_LambdaZ_CZ}
    \end{align}
    where $Z$ is the list index defined in \eqref{DefinitionZ}. The proofs of \eqref{Relation_LambdaX_LambdaZ}, \eqref{Relation_CX_CZ} and \eqref{Relation_LambdaZ_CZ} are presented in Appendix \ref{ProofTheoremConnection}.
\end{IEEEproof}
%%%%%
Theorem \ref{TheoremConnection} generalizes  \cite[Lemma 4]{Wu} derived by Wu and Joudeh in the error-free setting (i.e. $\epsilon = 0$), and \cite[Lemma 7]{SasonVerdu}  derived by Sason and Verdú in the error-free lossless setting (i.e. $\epsilon = D = 0$).
%%%%%
Combining the above theorem with the bounds on guessing moments derived in Section \ref{SectionSoftGuessingAlloingErrors} enables us to obtain bounds on the cumulant generating function as follows.
%%%%%
\begin{Corollary} 
\label{corollary:normalized_cumulant_bounds}
Given $\rho > 0$, $D \geq 0$, and $\epsilon \in [0,1)$, we have the upper bounds
\begin{align}
\label{eq:cumulant_upper_bound_1}
\Lambda_{X}^{\star} (\rho, D, \epsilon) \leq \frac{1}{\rho} \log \left( \exp \left(  \rho H^{\epsilon}_{\frac{1}{1+\rho}}(Z)\right) + \epsilon \right)
\end{align}
and
\begin{align}
\label{eq:cumulant_upper_bound_2}
\Lambda_{X}^{\star} (\rho, D, \epsilon) \leq \frac{1}{\rho} \log \left(1-\epsilon + 2^\rho \exp \left ( \rho H^{\epsilon}_{\frac{1}{1+\rho}}(X) - \rho \log \lfloor \exp(D) \rfloor \right ) + \epsilon \right). 
\end{align}
Moreover, we also have the lower bounds
\begin{align}
\Lambda_{X}^{\star} (\rho, D, \epsilon) & > \frac{1}{\rho} \log \left( 2^{-\rho}  (1+\log |\mathcal{X}|)^{-\rho} \exp \left( \rho H^{\epsilon}_{\frac{1}{1+\rho}}(Z) \right) + \epsilon \right) \\
& \geq \frac{1}{\rho} \log \left( 2^{-\rho} (1+\log |\mathcal{X}|)^{-\rho} \exp \left ( \rho H^{\epsilon}_{\frac{1}{1+\rho}}(X) - \rho \log \lfloor \exp(D) \rfloor \right ) + \epsilon \right).
\end{align}
\end{Corollary} 
%%%%%
The above corollary follows by combining Theorem \ref{TheoremConnection} with the bounds in Theorem \ref{MainTheorem_OneShot} and Proposition \ref{Proposition_OneShot}.
%%%%%
Note that Corollary \ref{corollary:normalized_cumulant_bounds} recovers the previous bounds in \cite[Theorem 13]{Shkel} and \cite[Remark 4]{Wu} as special cases.

Next, we consider the asymptotic regime where the source is an i.i.d. sequence $X^n$ and the $n$-letter log-loss distortion is given by \eqref{eq:n_letter_log_loss}.
%%%%%
In this case, we are interested in the per-symbol expected generalized length $\frac{1}{n}\Lambda_{X^n}^{\star} (\rho, D, \epsilon)$.
%%%%%
Similar to Proposition \ref{MainAsymptotic}, we assume that $V(X) < \infty$ and $T(X) < \infty$.
%%%%%
By combining the bounds in Theorem \ref{TheoremConnection} with the asymptotic result in Proposition \ref{MainAsymptotic}, we obtain the following asymptotic expansion. 
\begin{Corollary} 
\label{corollary:normalized_cumulant_asymptotic}
For any $\rho > 0$, $\epsilon \in (0,1)$, and $0 \leq D < H(X)$, it holds that
\begin{align}
   \frac{1}{n}\Lambda_{X^n}^{\star} (\rho, D, \epsilon) = H(X) - D - \sqrt{\frac{V(X)}{n}} \Phi^{-1}(\epsilon) + O\left(\frac{\log n}{n} \right).
\end{align}
\end{Corollary} 
%%%%%
It is clear from the above corollary that $  \frac{1}{n}\Lambda_{X^n}^{\star} (\rho, D, \epsilon)$ converges to the rate distortion function under log-loss as $n$ goes to infinity. We also remark that the above results and discussion can be extended to the case where side information is available to both the encoder and decoder, known as conditional source coding, by leveraging the results in Section \ref{section_Side_Information}.
%%%%%
\subsection{Error-Free Special Case}
\label{subsec:source_coding_error_free}
%%%%%
Our next goal is to shed light on how the two upper bounds in \eqref{eq:cumulant_upper_bound_1} and \eqref{eq:cumulant_upper_bound_2} compare.  
Recall from Proposition~\ref{proposition_upperComparison} and the preceding discussion that \eqref{eq:cumulant_upper_bound_1} is tighter than \eqref{eq:cumulant_upper_bound_2} for all $D$ such that $1 \leq \lfloor 2^D \rfloor \leq 2$.  
Beyond this regime, we are unable to establish a general relationship between the two bounds, so instead, we examine them numerically. To facilitate this, we focus on the error-free case $\epsilon = 0$. The following result follows directly from Corollary~\ref{corollary:normalized_cumulant_bounds} and Remark~\ref{RenyiEntropyandSmoothRenyiEntropy}.
%%%%%
\begin{Corollary} \label{CorollaryVariable-lengthUpperBound}
Given $\rho > 0$ and $D \geq 0$, we have
\begin{align}
\Lambda_{X}^{\star} (\rho, D, 0) \leq H_{\frac{1}{1+\rho}} \left( Z \right) \label{NewUpperBound}
\end{align}
and 
\begin{align}
\label{OldUpperBound}
\Lambda_{X}^{\star} (\rho, D, 0) \leq \frac{1}{\rho} \log \left(1 + 2^\rho \exp \left(\rho H_{\frac{1}{1+\rho}} (X) - \rho \log \lfloor \exp(D) \rfloor \right) \right). 
\end{align}
\end{Corollary} 
It is worth noting that while the upper bound in \eqref{OldUpperBound} was previously reported in \cite[Remark 4]{Wu}, the bound in \eqref{NewUpperBound}  is new; only a special case concerning the average length ($\rho = 0$) has appeared earlier in \cite[Theorem 13]{Shkel}.
%%%%
%\begin{IEEEproof}
%    From \eqref{MainUpperBound} and \eqref{ConnectionTwoFundamentalLimits}, we have
%    \begin{align}
%        \Lambda_{X}^{\star} (\rho, D, \epsilon) 
%        \leq \frac{1}{\rho} \log \left[ \exp \left ( \rho H^{\epsilon}_{\frac{1}{1+\rho}}(Z)\right ) \right]
%        = H^{\epsilon}_{\frac{1}{1+\rho}} \left( Z \right).
%    \end{align}
%    Hence, setting $\epsilon = 0$ and recalling Remark \ref{RenyiEntropyandSmoothRenyiEntropy}, we obtain \eqref{NewUpperBound}.
%\end{IEEEproof}
%%%%%%

\begin{figure}[t] 
\centering
\begin{minipage}[b]{0.32\columnwidth}
    \centering
    \includegraphics[width=1.0\columnwidth]{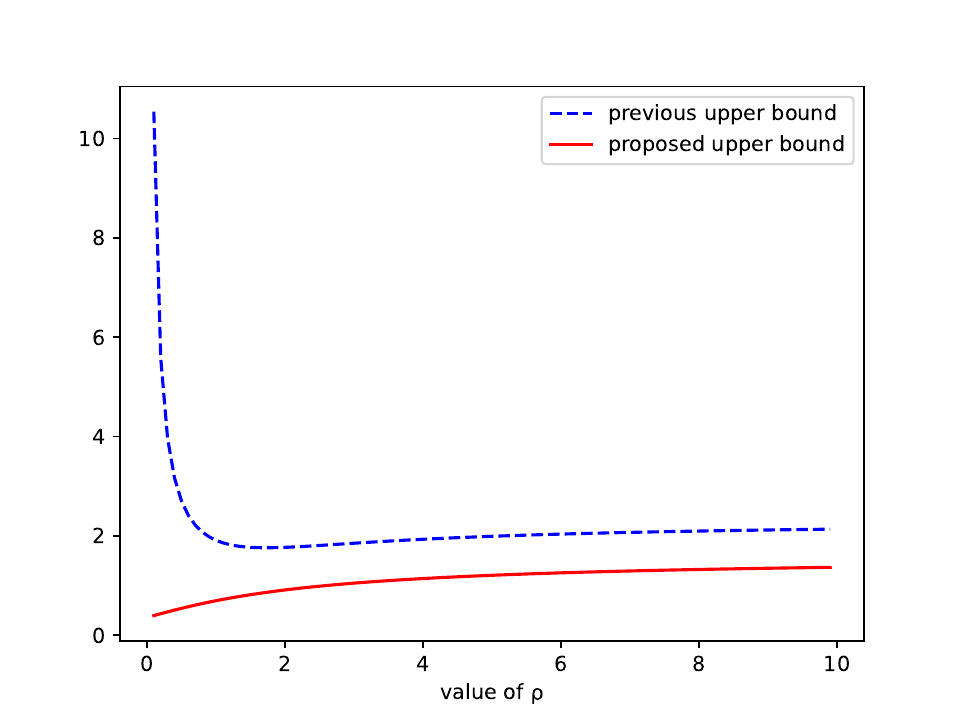}
\end{minipage}
\begin{minipage}[b]{0.32\columnwidth}
    \centering
    \includegraphics[width=1.0\columnwidth]{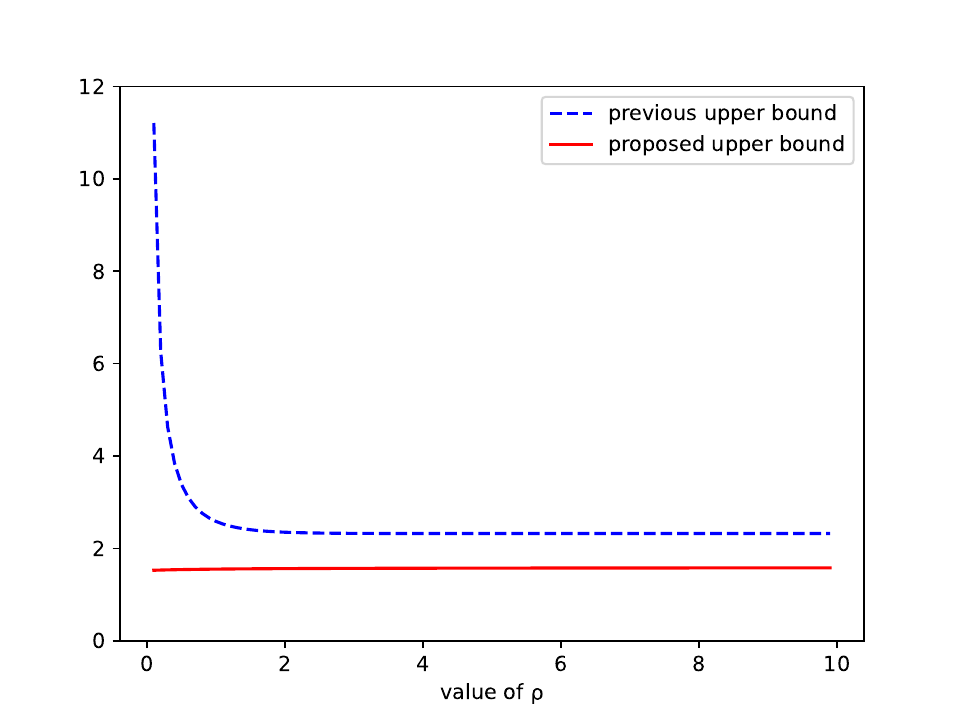}
\end{minipage}
\begin{minipage}[b]{0.32\columnwidth}
    \centering
    \includegraphics[width=1.0\columnwidth]{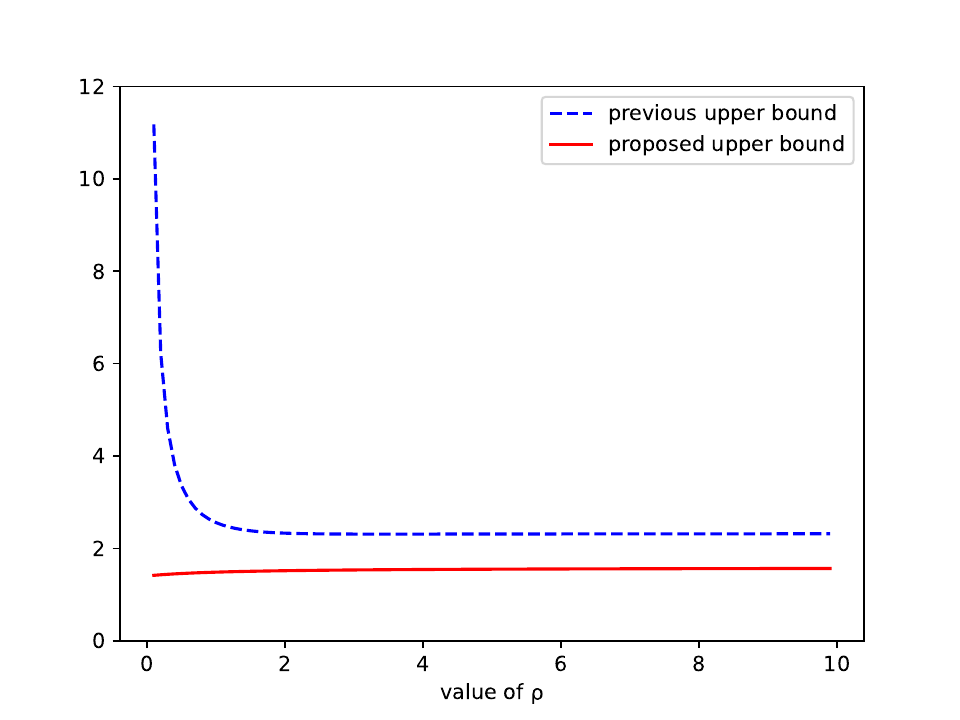}
\end{minipage}
\caption{Left: Case 1(a), Center: Case 1(b), Right: Case 1(c)}
\label{case1}
\end{figure}

\begin{figure}[t] 
\centering
\begin{minipage}[b]{0.32\columnwidth}
    \centering
    \includegraphics[width=1.0\columnwidth]{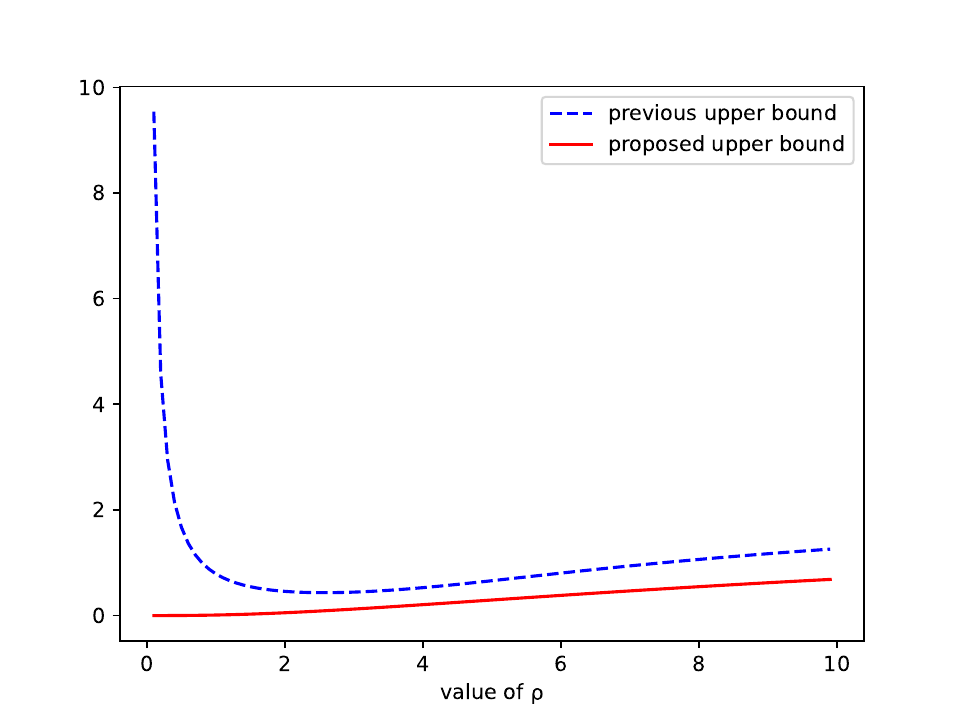}
\end{minipage}
\begin{minipage}[b]{0.32\columnwidth}
    \centering
    \includegraphics[width=1.0\columnwidth]{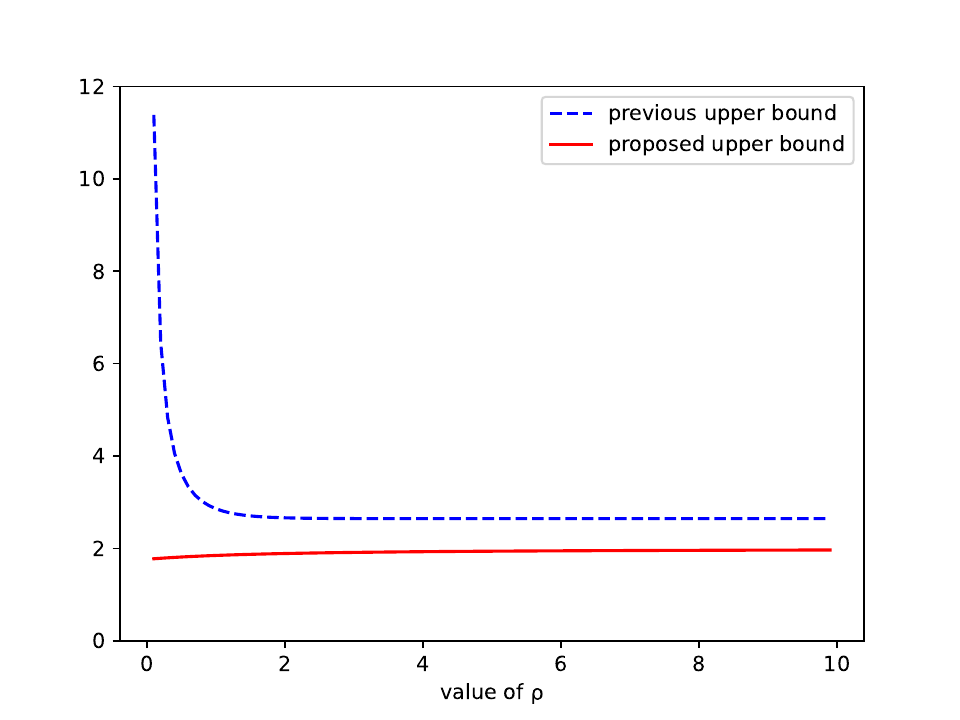}
\end{minipage}
\begin{minipage}[b]{0.32\columnwidth}
    \centering
    \includegraphics[width=1.0\columnwidth]{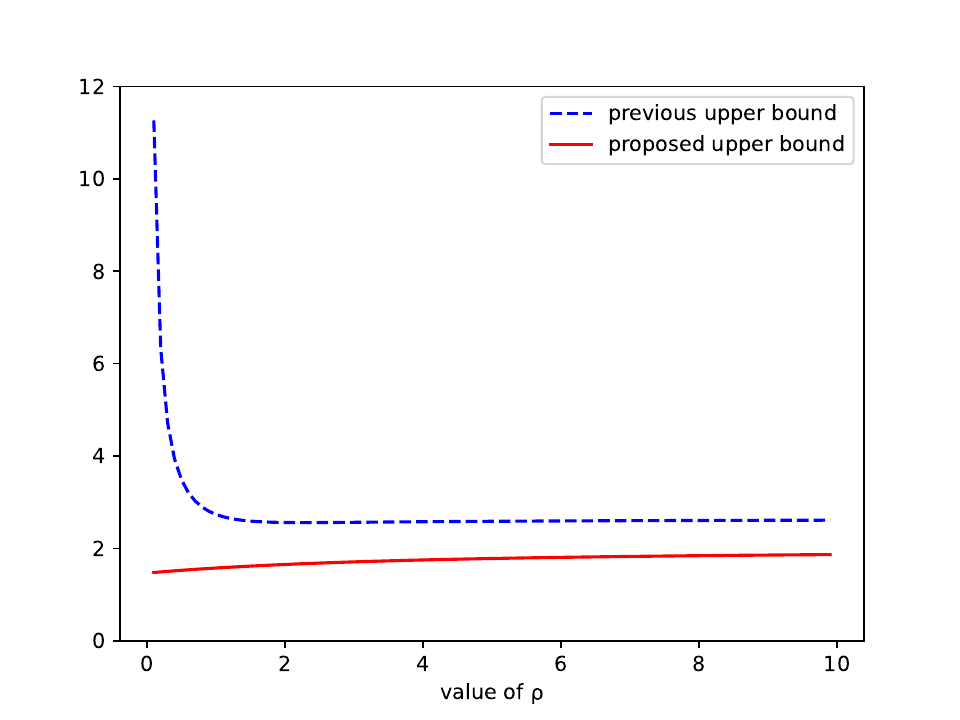}
\end{minipage}
\caption{Left: Case 2(a), Center: Case 2(b), Right: Case 2(c)}
\label{case2}
\end{figure}

As mentioned above, we know that \eqref{NewUpperBound} is tighter than \eqref{OldUpperBound} when $\lfloor 2^D \rfloor \leq 2$, therefore, in our numerical examples, we focus on the regime $\lfloor 2^D \rfloor > 2$. We consider the following examples.
%%%%%
\begin{itemize}
\item Case 1: We set $\mathcal{X}=\{1, 2, \ldots, 10 \}$, $D=2$, and $P_X$ as:
   \begin{enumerate}[label=(\alph*)]
   \item For $i=1, 2, \ldots, 9$, $P_X (i) = 1/2^{i}$ and $P_X (10) = 1/2^{9}$ (i.e., $P_X$ is a dyadic distribution\footnote{A probability distribution is called dyadic if each of the probabilities is equal to $2^{-k}$ for some integer $k$.}),
   \item For $i=1, 2, \ldots, 10$, $P_X (i) = 1/10$ (i.e., $P_X$ is a uniform distribution),
   \item $P_X$ is a randomly generated distribution.
   \end{enumerate}
\item Case 2: We set $\mathcal{X}=\{1, 2, \ldots, 50 \}$, $D=4$, and  $P_X$ as:
   \begin{enumerate}[label=(\alph*)]
   \item For $i=1, 2, \ldots, 49$, $P_X (i) = 1/2^{i}$ and $P_X (50) = 1/2^{49}$ (i.e., $P_X$ is a dyadic distribution),
   \item For $i=1, 2, \ldots, 50$, $P_X (i) = 1/50$ (i.e., $P_X$ is a uniform distribution),
   \item $P_X$ is a randomly generated distribution.
   \end{enumerate}
\end{itemize}
%%%%%%%%
For these settings, we plotted the previous upper bound in
\eqref{OldUpperBound} and the new proposed upper bound in \eqref{NewUpperBound} 
for $\rho$ values between $0.1$ and $10$.
The result for Case 1 is shown in Fig.\ \ref{case1} and the result for Case 2 is shown in Fig.\ \ref{case2}.
%%%%%%
These figures suggest that the proposed upper bound in \eqref{NewUpperBound} is tighter than the previous upper bound in \eqref{OldUpperBound}.
%%%%%%
\begin{Remark} \label{UpperBoundRemark}
We observe from   Fig.\ \ref{case1} and  Fig.\ \ref{case2} that the previous upper bound in \eqref{OldUpperBound} diverges as $\rho \to 0$. This is seen from
\begin{align}
& \lim_{\rho \to 0} \frac{\log \left[1 + 2^\rho \exp \left(\rho H_{\frac{1}{1+\rho}} (X) - \rho \log \lfloor \exp(D) \rfloor \right) \right]}{\rho}  = \frac{\log2}{\lim_{\rho \to 0} \rho} = \infty.
\end{align}
%\begin{align}
%& \lim_{\rho \to 0} \frac{\log \left[1 + 2^\rho \exp \left(\rho H_{\frac{1}{1+\rho}} (X) - \rho \log %\lfloor \exp(D) \rfloor \right) \right]}{\rho} \\
%& \quad = \frac{\lim_{\rho \to 0} \log \left[1 + 2^\rho \exp \left(\rho H_{\frac{1}{1+\rho}} (X) - \rho %\log \lfloor \exp(D) \rfloor \right) \right]}{\lim_{\rho \to 0} \rho} \\
%& \quad = \frac{\log2}{\lim_{\rho \to 0} \rho}.
%\end{align}
\end{Remark}

\section{Concluding remarks} \label{SectionConclusion}
%%%%%
We formulated and studied the problem of soft guessing under log-loss while allowing errors.
%%%%%
We identified the optimal guessing strategy in this setting and derived single-shot upper and lower bounds on the guessing moments, expressed in terms of smooth Rényi entropies and the distortion level $D$. These bounds recover previous bounds, derived for special cases of the problem.
%%%%%
We further derived asymptotic expansions for i.i.d. sources, leading to a sharp characterization of the guessing exponents.
%%%%%
The results were also extended to the case where side information is available to the guesser.
%%%%%
Finally, we established a relationship between guessing moments and the normalized cumulant generating function of codeword lengths in a corresponding variable-length source coding problem. This enabled us to bound the latter using the bounds obtained for the former.
%%%%%

Potential directions for future work include clarifying whether the relationship in \eqref{eq:upperComparison}, which we verified for all $D$ such that $\lfloor 2^D \rfloor \leq 2$, continues to hold when $\lfloor 2^D \rfloor > 2$.
%%%%%
The numerical results we reported above suggest that this may be the case, but currently we have no formal proof.
%%%%%
Another direction is to refine the asymptotic analysis of the guessing moment under side information, i.e., $M^\star_{X^n|Y^n}(\rho, D, \epsilon)$. While we established expansions for $M^\star_{X^n}(\rho, D, \epsilon)$ up to the first- and second-order terms, for $M^\star_{X^n|Y^n}(\rho, D, \epsilon)$, we obtained only the first-order term. 
%%%%%
This limitation stems from the results of Sakai and Tan \cite{Sakai}, where the smooth R\'enyi entropy was expanded up to the third-order term (see Lemma \ref{Lemma_Asymptotic}), but the conditional smooth R\'enyi entropy only up to the first-order term (see Lemma \ref{Lemma_Asymptotic_SideInformation}).
%%%%%
Extending the expansion of the conditional smooth R\'enyi entropy, and thereby establishing the second-order term of $M^\star_{X^n|Y^n}(\rho, D, \epsilon)$, remains an open problem.

%%%%%%
%%%%%%
\appendices
%%%%%%
\section{Proof of Lemma \ref{Lemma_chain_rule_conditional_smooth_Renyi}} \label{proof_Lemma_chain_rule_conditional_smooth_Renyi}
%%%%%%
%%%%%%
From the explicit formulas in Lemma \ref{Lemma_explicit_Koga} and Lemma \ref{Lemma_explicit_Kuzuoka}, we can write
\begin{align}
\label{eq:cond_smooth_ren_entropy_formula}
    H^{\epsilon}_{\alpha}(X, Z|Y) = \inf_{(\epsilon_y) \in \mathcal{E}_{0}(\epsilon)} \frac{\alpha}{1 - \alpha} \log \left( \sum_{y \in \mathcal{Y}} P_Y(y) \exp \left( \frac{1 - \alpha}{\alpha} H_{\alpha}^{\epsilon_y} \left( P_{X,Z|Y=y}\right) \right) \right ), 
\end{align}
%%%%
where $H_{\alpha}^{\epsilon_y} \left( P_{X,Z|Y=y}\right)$ is the smooth R\'enyi entropy of $P_{X,Z|Y=y}$.\footnote{Setting $\epsilon = 0$ in \eqref{eq:cond_smooth_ren_entropy_formula} recovers a known identity connecting the Arimoto-R\'enyi conditional entropy to R\'enyi entropies of constituent conditional pmfs.} 
%%%%
From Corollary \ref{Corollary_smooth_chain_rule}, we know that for every $y \in \mathcal{Y}$, we have 
\begin{align}
    H_{\alpha}^{\epsilon_y} \left( P_{X,Z|Y=y} \right) \leq \tilde{H}_{\alpha}^{0} \left( P_{X|Z,Y=y} \right) + 
    H_{\alpha}^{\epsilon_y} \left( P_{Z|Y=y}\right),
\end{align}
where $\tilde{H}^{0}_{\alpha}(P_{X|Z,Y=y})$ is bounded above as
\begin{align}
\tilde{H}^{0}_{\alpha}(P_{X|Z,Y=y}) & = \max_{z:P_{Z|Y}(z|y) > 0} H_{\alpha}(P_{X|Z=z,Y=y}) \\
& \leq \max_{y:P_{Y}(y) > 0} \max_{z:P_{Z|Y}(z|y) > 0} H_{\alpha}(P_{X|Z=z,Y=y}) \\
& = \tilde{H}^{0}_{\alpha}(X|Z,Y).
\end{align}
Plugging everything back into \eqref{eq:cond_smooth_ren_entropy_formula}, we get 
\begin{align}
    H^{\epsilon}_{\alpha}(X, Z|Y) & \leq \inf_{(\epsilon_y) \in \mathcal{E}_{0}(\epsilon)} \frac{\alpha}{1 - \alpha} \log \left( \sum_{y \in \mathcal{Y}} P_Y(y) \exp \left( \frac{1 - \alpha}{\alpha} \left( \tilde{H}^{0}_{\alpha}(X|Z,Y) + H_{\alpha}^{\epsilon_y} \left( P_{Z|Y=y}\right) \right) \right) \right ) \\
    & = \tilde{H}^{0}_{\alpha}(X|Z,Y) + \inf_{(\epsilon_y) \in \mathcal{E}_{0}(\epsilon)} \frac{\alpha}{1 - \alpha} \log \left( \sum_{y \in \mathcal{Y}} P_Y(y) \exp \left( \frac{1 - \alpha}{\alpha} H_{\alpha}^{\epsilon_y} \left( P_{Z|Y=y}\right) \right) \right) \\
    & = \tilde{H}^{0}_{\alpha}(X|Z,Y) + H_{\alpha}^{\epsilon}(Z|Y),
\end{align}
which completes the proof of Lemma \ref{Lemma_chain_rule_conditional_smooth_Renyi}.
%%%%%%%
%%%%%%%
\section{Proof of Proposition \ref{PropositionOptimalStrategy}} \label{ProofOptimalStrategy}
Consider an arbitrary $D$-admissible strategy $(\mathcal{G}_{\mathrm{soft}}(D),\pi)$ that satisfies the error probability constraint 
\begin{align}
    P_e = \mathbb{P}[G(X)=0] \leq \epsilon. \label{ErrorProbabilityConstraint}
\end{align}
To show Proposition \ref{PropositionOptimalStrategy}, it suffices to prove that
\begin{align}
    \mathbb{E}\left[G^\star(X)^{\rho} \right] \leq \mathbb{E}\left[G(X)^{\rho} \right]. \label{GuessingMomentInequality}
\end{align}
To this end, we follow in the footsteps of Sakai and Tan in their proof of \cite[Lemma 10]{Sakai}, which in turn, employs the competitive optimality idea in the proof of Ar\i kan \cite{Arikan}.
%%%%%
In particular, to prove \eqref{GuessingMomentInequality}, we will show that
\begin{align}
    \mathbb{P}\left[G^\star(X) \leq k \right] \geq \mathbb{P}\left[G(X) \leq k \right] \quad (\forall k \in \{0, 1, \ldots, N \}). \label{GuessingProbabilityInequality}
\end{align}
This is because if we can prove \eqref{GuessingProbabilityInequality}, then it holds that
\begin{align}
    \mathbb{E}\left[G^\star(X)^{\rho} \right]
    &= \sum_{k=1}^{N} \big(k^\rho - (k-1)^\rho \big) \mathbb{P}\left[G^\star(X) \geq k \right] \\
    &\leq \sum_{k=1}^{N} \big(k^\rho - (k-1)^\rho \big) \mathbb{P}\left[G(X) \geq k \right] \\
    &=\mathbb{E}\left[G(X)^{\rho} \right].
\end{align}
We proceed by noting that for every $k = 0, 1, 2, \ldots, K-1$, where $K$ is as defined in \eqref{Definition_K}, we have
\begin{align}
    \mathbb{P}\left[G(X) \leq k \right]
    &= \mathbb{P}\left[G(X) = 0 \right] + \sum_{i=1}^{k} \mathbb{P}\left[G(X) = i \right] \label{appendix_probability_upper_bound_optimal_strategy_1} \\
    &= \mathbb{P}\left[G(X) = 0 \right] + \sum_{i=1}^{k} \lambda_i \left(\sum_{x \in g^{-1}(i)} P_X(x) \right) \\
    &\leq \mathbb{P}\left[G^\star(X) = 0 \right] + \sum_{i=1}^{k} \lambda_i \left(\sum_{x \in g^{-1}(i)} P_X(x) \right) \label{appendix_probability_upper_bound_optimal_strategy_2} \\
    &\leq \mathbb{P}\left[G^\star(X) = 0 \right] + \sum_{i=1}^{k} \lambda^\star_i \left(\sum_{x \in (g^\star)^{-1}(i)} P_X(x) \right) \label{appendix_probability_upper_bound_optimal_strategy_3} \\
    &= \mathbb{P}\left[G^\star(X) = 0 \right] + \sum_{i=1}^{k} \mathbb{P}\left[G^\star(X) = i \right] \\
    &= \mathbb{P}\left[G^\star(X) \leq k \right],
\end{align}
where \eqref{appendix_probability_upper_bound_optimal_strategy_2} is due to the inequality $\mathbb{P}\left[G(X) = 0 \right] \leq \mathbb{P}\left[G^\star(X) = 0 \right]$ (see \eqref{OptimalErrorProbability} and \eqref{ErrorProbabilityConstraint}); and \eqref{appendix_probability_upper_bound_optimal_strategy_3} follows from
\begin{align}
    \sum_{i=1}^{k} \lambda_i \left(\sum_{x \in g^{-1}(i)} P_X(x) \right) \leq \sum_{i=1}^{k} \lambda^\star_i \left(\sum_{x \in (g^\star)^{-1}(i)} P_X(x) \right). \label{appendix_probability_upper_bound_optimal_strategy_4}
\end{align}
To verify \eqref{appendix_probability_upper_bound_optimal_strategy_4}, the following lemma by Shkel and Verd\'u \cite{Shkel} is useful.
\begin{Lemma}  [{\cite[Lemma 1]{Shkel}}] \label{Lemma_D_cover}
    For the log-loss defined in \eqref{Def_LogLoss}, we say that $\hat{P} \in \mathcal{P}(\mathcal{X})$ $D$-covers $x \in \mathcal{X}$ if $d(x, \hat{P}) \leq D$. Then, no soft reconstruction $\hat{P}$ can $D$-cover more than $\lfloor \exp(D) \rfloor$ realization of $X$, i.e., it holds that
    \begin{align}
        \big| \big\{x \in \mathcal{X} : \hat{P} ~ D\text{-covers}~x \big\} \big| \leq \lfloor \exp(D) \rfloor,  \ \ \text{for every} \ \hat{P} \in \mathcal{P}(\mathcal{X}).
    \end{align}
\end{Lemma}
Recall that in the scheme $\mathcal{G}_{\mathrm{soft}}^\star(D)$, each of the lists $\mathcal{L}_1, \mathcal{L}_2, \ldots, \mathcal{L}_{K-1}$ contains $\lfloor \exp(D) \rfloor$ realizations of $X$ (see \eqref{List}), and  higher probability realizations are assigned shorter guessing orders.
%%%%%
Combining this with Lemma \ref{Lemma_D_cover}, we get 
\begin{align}
    \sum_{x \in g^{-1}(i)} P_X(x) \leq \sum_{x \in (g^\star)^{-1}(i)} P_X(x) \label{appendix_probability_upper_bound_optimal_strategy_5}
\end{align}
for every $i = 1, 2, \ldots, K-1$. Moreover, from the property of $\pi^\star$ that $\pi^\star_i = 0$ for $i=1, 2, \ldots, K-1$ (see \eqref{OptimalGiveUpProbability}), and hence $\lambda^\star_i = 1$ for $i=1, 2, \ldots, K-1$ (see \eqref{OptimalLambda}), we get
\begin{align}
    \lambda_i \leq \lambda^\star_i \label{appendix_probability_upper_bound_optimal_strategy_6}
\end{align}
for every $i = 1, 2, \ldots, K-1$. Therefore, by combining \eqref{appendix_probability_upper_bound_optimal_strategy_5} and \eqref{appendix_probability_upper_bound_optimal_strategy_6}, we have \eqref{appendix_probability_upper_bound_optimal_strategy_4}.
%%%%%

To complete the proof, we wish to show that for every $k = K, K+1, \ldots, N$, it also holds that
\begin{align}
    \mathbb{P}\left[G(X) \leq k \right] \leq \mathbb{P}\left[G^\star(X) \leq k \right].
\end{align}
This follows by noting that 
%%%%%
\begin{align}
    \mathbb{P}\left[G^\star(X) \leq K \right]
    &= \mathbb{P}\left[G^\star(X) = 0 \right] + \sum_{i=1}^{K} \mathbb{P}\left[G^\star(X) = i \right] \\
    &= \epsilon + \sum_{i=1}^{K-1} \left(\sum_{x \in (g^\star)^{-1}(i)} P_X(x) \right) + \lambda^\star_K \left(\sum_{x \in (g^\star)^{-1}(K)} P_X(x) \right) \\
    &= \epsilon + \sum_{i=1}^{i^*_X} Q^\epsilon_X(i) \\
    &= \epsilon + (1-\epsilon) \\
    &= 1.
\end{align}
Therefore, we have shown that \eqref{GuessingProbabilityInequality} holds, which completes the proof.
%%%%%%%
\section{Proof of Proposition \ref{Proposition_OneShot}}
\label{Appendix:Proof_proposition_explicit_bounds}
%%%%%
\subsection{Upper Bound} \label{ProofAchievability}
As in Section \ref{SectionOptimalGuessing}, let $L:= \lfloor \exp(D) \rfloor$. First, we consider the case where $L > 1$. From \eqref{AchievabilityStep(b)}, we have
\begin{align}
    M^\star_X (\rho, D, \epsilon)
    &=\sum_{i=1}^{K-1} \left( \sum_{j=(i-1) L+1}^{iL}  Q^\epsilon_X(j) \right) i^{\rho} + \left( \sum_{i=(K-1) L+1}^{i^*_X} Q^\epsilon_X(i) \right) K^{\rho} \\
    &= \sum_{j=1}^{i^*_X} Q^\epsilon_X(j) \left \lceil \frac{j}{L} \right \rceil^\rho \label{AppendixAchievability} \\
    &= \sum_{j=1}^{i^*_X} Q^\epsilon_X(j) \left \lceil \frac{1}{L} \sum_{k : k \leq j} 1 \right \rceil^\rho \\
    &\leq \sum_{j=1}^{i^*_X} Q^\epsilon_X(j) \left \lceil \frac{1}{L} \sum_{k : k \leq j} \left( \frac{Q^\epsilon_X(k)}{Q^\epsilon_X(j)} \right)^{\frac{1}{1+\rho}} \right \rceil^\rho \label{appendix_achievability_stepA} \\
    &\leq \sum_{j=1}^{i^*_X} Q^\epsilon_X(j) \left \lceil \frac{1}{L} \sum_{k=1}^{i^*_X} \left( \frac{Q^\epsilon_X(k)}{Q^\epsilon_X(j)} \right)^{\frac{1}{1+\rho}} \right \rceil^\rho \\
    &\leq \sum_{j=1}^{i^*_X} Q^\epsilon_X(j) \left \{1+2^\rho \left( \frac{1}{L} \sum_{k=1}^{i^*_X} \left( \frac{Q^\epsilon_X(k)}{Q^\epsilon_X(j)} \right)^{\frac{1}{1+\rho}} \right)^\rho \right \} \label{appendix_achievability_stepB} \\
    &= \sum_{j=1}^{i^*_X} Q^\epsilon_X(j) + \left( \frac{2}{L} \right)^\rho \left (\sum_{j=1}^{i^*_X} [Q^\epsilon_X(j)]^{\frac{1}{1+\rho}} \right )^{1+\rho} \\
    &= 1-\epsilon + \left( \frac{2}{L} \right)^\rho \exp \left ( \rho H^{\epsilon}_{\frac{1}{1+\rho}}(X) \right ) \label{appendix_achievability_stepC} \\
    &= 1-\epsilon + 2^\rho \exp \left ( \rho H^{\epsilon}_{\frac{1}{1+\rho}}(X) - \rho \log \lfloor \exp(D) \rfloor \right ), 
\end{align}
where \eqref{appendix_achievability_stepA} follows from the definition of $Q^\epsilon_X$ in Lemma \ref{Lemma_explicit_Koga};
\eqref{appendix_achievability_stepB} is due to the of Bunte-Lapidoth inequality $\lceil \xi \rceil^\rho < 1 + 2^\rho \xi^\rho $ for $\xi \geq 0$ and $\rho > 0$  (see \cite[Eq.\ (26)]{Bunte}); and \eqref{appendix_achievability_stepC} follows from Lemma \ref{Lemma_explicit_Koga}.

Next, we consider the case where $L = 1$. Here, instead of \eqref{AppendixAchievability}, we have
\begin{align}
    M^\star_X (\rho, D, \epsilon) = \sum_{j=1}^{i^*_X}  Q^\epsilon_X(j) j^\rho,
\end{align}
and hence we do not need to use the inequality $\lceil \xi \rceil^\rho < 1 + 2^\rho \xi^\rho $ as in the above case. Hence, we directly obtain
\begin{align}
    M^\star_X (\rho, D, \epsilon) \leq \exp \left ( \rho H^{\epsilon}_{\frac{1}{1+\rho}}(X) \right ).
\end{align}
%%%%%%%
\subsection{Lower Bound} \label{ProofConverse}
%%%%%%%
We utilize the properties of the smooth R\'enyi entropy. Since $\rho > 0$, it holds that $0 < 1/(1+\rho) <1$. Hence, we have
\begin{align}
     H^{\epsilon}_{\frac{1}{1+\rho}}(Z)
     & \geq H^{\epsilon}_{\frac{1}{1+\rho}}(X, Z) -  \tilde{H}^{0}_{\frac{1}{1+\rho}}(X|Z) \label{appendix_converse_stepA} \\
     & \geq H^{\epsilon}_{\frac{1}{1+\rho}}(X) -  \tilde{H}^{0}_{\frac{1}{1+\rho}}(X|Z), \label{appendix_converse_stepB}
\end{align}
where $Z$ is defined by \eqref{DefinitionZ}; \eqref{appendix_converse_stepA} follows from the chain rule in Corollary \ref{Corollary_smooth_chain_rule}; and \eqref{appendix_converse_stepB} is due to monotonicity in Lemma \ref{Lemma_smooth_monotonic}.
Next, we bound the term $\tilde{H}^{0}_{\frac{1}{1+\rho}}(X|Z)$ above as
\begin{align}
    \tilde{H}^{0}_{\frac{1}{1+\rho}}(X|Z) 
    & = \max_{z:P_Z(z) > 0} H_{\frac{1}{1+\rho}}(P_{X|Z=z}) \\
    & \leq \max_{z:P_Z(z) > 0} \log|\mathrm{supp}(P_{X|Z=z})|.
\end{align}
Since $Z$ represents the index of the list containing $X$, as defined in \eqref{DefinitionZ}, then the cardinality of the support of $X$ given $Z = z$ is at most $\lfloor \exp(D) \rfloor$. Hence, for every $z$ with $P_Z(z) > 0$, we have
\begin{align}
    \log |\mathrm{supp}(P_{X|Z=z})| \leq \log \lfloor \exp(D) \rfloor,
\end{align}
from which it follows that
\begin{align}
    \tilde{H}^{0}_{\frac{1}{1+\rho}}(X|Z) \leq \log \lfloor \exp(D) \rfloor. \label{upper_bound_Tilde_H}
\end{align}
Finally, combining \eqref{MainLowerBound}, \eqref{appendix_converse_stepB}, and \eqref{upper_bound_Tilde_H}, we obtain \eqref{Converse}.
%%%%%
%%%%%
\section{Proof of Proposition \ref{OneShotUpperBoundSideInformation}} \label{Appendix:Proof_proposition_explicit_bounds_side_inf}

\subsection{Upper Bound}
By the same argument used in the upper bound proof for Proposition \ref{Proposition_OneShot}, for any given side information realization $y$, $M_{X|Y = y}^{\star} (\rho, D, \epsilon_y)$ on the right-hand side of \eqref{relation_optimal_moments_side_information} is upper bounded  as 
\begin{align}
    M_{X|Y = y}^{\star} (\rho, D, \epsilon_y) \leq 1-\epsilon_y + \left( \frac{2}{L} \right)^\rho  \left (\sum_{j=1}^{i_{X|y}^*} [Q^{\epsilon_y}_{X|Y}(x^j_y | y)]^{\frac{1}{1+\rho}} \right )^{1+\rho}.
\end{align}
Next, we take the expectation with respect to $Y$ and then the infimum over $(\epsilon_y)\in \mathcal{E}_0(\epsilon )$. From the explicit form of Kuzuoka's conditional smooth R\'enyi entropy in Lemma \ref{Lemma_explicit_Kuzuoka}, we obtain \eqref{explicit_upper_side_information}. The upper bound for $ 0 \leq D < 1$ is similarly proved.
%%%%%%%
\subsection{Lower Bound}
From  the conditional chain rule in Lemma \ref{Lemma_chain_rule_conditional_smooth_Renyi} and the conditional monotonicity in Lemma \ref{Lemma_monotonicity_conditional_smooth_Renyi}, we obtain
\begin{align}
     H^{\epsilon}_{\frac{1}{1+\rho}}(Z|Y)
     & \geq H^{\epsilon}_{\frac{1}{1+\rho}}(X, Z | Y) -  \tilde{H}^{0}_{\frac{1}{1+\rho}}(X|Z, Y) \\
     & \geq H^{\epsilon}_{\frac{1}{1+\rho}}(X | Y) -  \tilde{H}^{0}_{\frac{1}{1+\rho}}(X|Z, Y). \label{explicit_lower_boud_side_information_stepA}
\end{align}
Using similar argument to the ones used in the lower bound proof for Proposition \ref{Proposition_OneShot},
the term $\tilde{H}^{0}_{\frac{1}{1+\rho}}(X|Z,Y)$ is bounded as
\begin{align}
    \tilde{H}^{0}_{\frac{1}{1+\rho}}(X|Z,Y) 
    & = \max_{y:P_{Y}(y) > 0} \max_{z:P_{Z|Y}(z|y) > 0} H_{\alpha}(P_{X|Z=z,Y=y}) \\
    & \leq \max_{y:P_{Y}(y) > 0} \max_{z:P_{Z|Y}(z|y) > 0} \log|\mathrm{supp}(P_{X|Z=z,Y=y})| \\
    & \leq \log \lfloor \exp(D) \rfloor. \label{explicit_lower_boud_side_information_stepB}
\end{align}
Finally, combining \eqref{MainLowerBound_side}, \eqref{explicit_lower_boud_side_information_stepA}, and \eqref{explicit_lower_boud_side_information_stepB}, we obtain \eqref{explicit_lower_side_information}.
%%%%%%
%%%%%%
\section{Proof of Theorem \ref{TheoremConnection}} \label{ProofTheoremConnection}
%%%%%%
Here we prove \eqref{Relation_LambdaX_LambdaZ}, \eqref{Relation_CX_CZ} and \eqref{Relation_LambdaZ_CZ}.
%%%%%
We begin by finding an optimal variable-length source code that attains $\Lambda_{X}^{\star}(\rho, D, \epsilon)$ in \eqref{FundamentalLimitLossySourceCode}. We denote this by $(f^\star_X, \varphi^\star_X)$.
%%%%%
Note that in \cite[Section IV]{Shkel}, Shkel and Verd\'u found a competitively optimal variable-length source code under log-loss with no errors.
%%%%%%
The code we present next can be seen as an extension of their code.
%%%%%%

Let $L:= \lfloor \exp(D) \rfloor$. 
For $l=1, 2, \ldots, \left \lceil \frac{|\mathcal{X}|}{L} \right \rceil-1$, define the following sets (or lists)
\begin{align}
    \mathcal{L}_l := \{(l-1) L+1,  (l-1) L +2, \ldots, l L \} \label{SourceCodeList1}
\end{align}
and
\begin{align}
    \mathcal{L}_{\left \lceil \frac{|\mathcal{X}|}{L} \right \rceil} := \left \{\left(\left \lceil \frac{|\mathcal{X}|}{L} \right \rceil-1 \right) L+1, \ldots, |\mathcal{X}| \right \}. \label{SourceCodeList2}
\end{align}
Let $\mathcal{L}_{l^*} \in \left \{\mathcal{L}_1, \mathcal{L}_2, \ldots, \mathcal{L}_{\left \lceil \frac{|\mathcal{X}|}{L} \right \rceil} \right \}$ be the set such that $i^*_X \in \mathcal{L}_{l^*}$, where $i^*_X$ is defined in \eqref{definition_i_star}.
%%%%
The encoder $f^*_X$ maps the elements in $\mathcal{L}_1, \mathcal{L}_2, \ldots, \mathcal{L}_{l^* -1}$ to the elements of $\{0,1\}^*$ in the lexicographic order, i.e., 
\begin{align}
    f^\star_X(x) = 
    \begin{cases}
        \varnothing, & x \in \mathcal{L}_1, \\
        0, & x \in \mathcal{L}_2, \\
        1, & x \in \mathcal{L}_3, \\
        00, & x \in \mathcal{L}_4, \\
        01, & x \in \mathcal{L}_5, \\
        10, & x \in \mathcal{L}_6, \\
        11, & x \in \mathcal{L}_7, \\
        000, & x \in \mathcal{L}_8, \\
        & \vdots \\
        {\sf s}, & x \in \mathcal{L}_{l^* -1},
    \end{cases}
\end{align}
where ${\sf s} \in \{0,1\}^*$ is a binary string of length $\lfloor \log (l^* -1) \rfloor$. Second, $f^\star_X$ maps each element $x \in \mathcal{L}_{l^*}$ as
\begin{align}
    f^\star_X(x) = 
    \begin{cases}
        \varnothing, & \mathrm{with ~ probability} ~ \alpha, \\
        {\sf s}^+, &\mathrm{with ~ probability} ~ 1-\alpha,
    \end{cases}
\end{align}
where ${\sf s}^+ \in \{0,1\}^*$ is a binary string that follows ${\sf s}$ in the lexicographic order, and $\alpha$ is chosen so that
\begin{align}
\label{eq:error_code_X}
    \epsilon = \alpha \sum_{x \in \mathcal{L}_{l^*}} P_X(x) + \sum_{i=l^* +1}^{\left \lceil \frac{|\mathcal{X}|}{L} \right \rceil} \sum_{x \in \mathcal{L}_{i}} P_X(x).
\end{align}
Finally, $f^\star_X$ maps all the elements in $\mathcal{L}_{l^* +1}, \mathcal{L}_{l^* +2}, \ldots, \mathcal{L}_{\left \lceil \frac{|\mathcal{X}|}{L} \right \rceil}$ to the empty string $\varnothing$.
The decoder $\varphi^\star_X$ is defined as
\begin{align}
    \varphi^\star_X(s) = 
    \begin{cases}
        \hat{P}_1, & s = \varnothing, \\
        \hat{P}_2, & s = 0, \\
        \hat{P}_3, & s = 1, \\
        \hat{P}_4, & s = 00, \\
        & \vdots \\
        \hat{P}_{l^* - 1}, & s =  {\sf s}, \\
        \hat{P}_{l^*}, & s =  {\sf s}^+,
    \end{cases}
\end{align}
where
\begin{align}
    \hat{P}_i (x):= 
    \begin{cases}
        \frac{1}{|\mathcal{L}_i|}, & \forall x \in \mathcal{L}_i, \\
        0, & \mathrm{otherwise}.
    \end{cases}
\end{align}

From the code construction, it can be verified that excess distortion probability satisfies
\begin{align}
\mathbb{P}[d(X, \varphi^\star_X (f^\star_X (X))) > D] = \epsilon, 
\end{align}
and thus $(f^\star_X, \varphi^\star_X)$ is a $(\Lambda, D, \rho, \epsilon)$-code for some $\Lambda$. We see that $(f^\star_X, \varphi^\star_X)$ is optimal because no codeword can $D$-cover more than $L=\lfloor \exp(D) \rfloor$ elements in $\mathcal{X}$ (see Lemma \ref{Lemma_D_cover}), and $f^\star_X$ assigns shorter strings to the more likely elements. Therefore
\begin{align}
    \Lambda_{X}^{\star} (\rho, D, \epsilon) = \frac{1}{\rho} \log \mathbb{E}[\exp \{\rho \ell(f^\star_X(X)) \}].
\end{align}
%%%%%%%

%%%%%%%
Next, to prove \eqref{Relation_LambdaX_LambdaZ}, we look at $\Lambda_{Z}^{\star} (\rho, 0, \epsilon)$, where $Z$ is defined in \eqref{DefinitionZ}. We repeat its definition here for convenience.
\begin{align}
    Z := \left \lceil \frac{X}{\lfloor \exp(D) \rfloor} \right \rceil. 
\end{align}
%%%%%%
For coding $Z$, we have $D=0$ and the condition $\mathbb{P}[d(Z, \varphi (f (Z))) > D] \leq \epsilon$ becomes
\begin{align}
    \mathbb{P}[Z~ \text{is correctly recovered}] \geq 1 - \epsilon, \label{CorrectProbability}
\end{align}
because $d(z, \hat{P})=0$ if and only if $\hat{P}(z)=1$, i.e., $\hat{P}$ is a correct ``hard'' reconstruction. 
%%%%%%
An optimal code that attains $\Lambda_{Z}^{\star} (\rho, 0, \epsilon)$, denoted by $(f^\star_Z, \varphi^\star_Z)$, is obtained by setting the encoder as $f^\star_Z(z) = f^\star_X( (z-1)L + 1 )$ and the decoder as
\begin{align}
    \varphi^\star_Z(s) = 
    \begin{cases}
        \hat{P}_1, & s = \varnothing, \\
        \hat{P}_2, & s = 0, \\
        \hat{P}_3, & s = 1, \\
        \hat{P}_4, & s = 00, \\
        & \vdots \\
        \hat{P}_{l^* - 1}, & s =  {\sf s}, \\
        \hat{P}_{l^*}, & s =  {\sf s}^+,
    \end{cases}
\end{align}
where  
\begin{align}
    \hat{P}_i (z):= 
    \begin{cases}
        1, & z=i, \\
        0, & \mathrm{otherwise}.
    \end{cases}
\end{align}
%%%%%%
From the choice of the encoder $f^\star_Z$ and \eqref{eq:error_code_X}, it immediately follows that 
\begin{align}
    \epsilon = \alpha P_Z(l^*) + \sum_{i=l^* +1}^{\left \lceil \frac{|\mathcal{X}|}{L} \right \rceil} P_Z(i). \label{AppendixE_choice_alpha}
\end{align}
%%%%%%
The code $(f^\star_Z, \varphi^\star_Z)$ coincides with the optimal code of Kostina \emph{et al.} in \cite[Section II]{Kostina}.
%%%%%%
Moreover, $Z$ corresponds to the index of the lists in \eqref{SourceCodeList1}, which explains the close relationship between $(f^\star_X, \varphi^\star_X)$ and $(f^\star_Z, \varphi^\star_Z)$.
%%%%%
It follows that  \eqref{Relation_LambdaX_LambdaZ} is obtained as
\begin{align}
    \Lambda_{X}^{\star} (\rho, D, \epsilon) 
    &= \frac{1}{\rho} \log \mathbb{E}[\exp \{\rho \ell(f^\star_X(X)) \}] \\
    &= \frac{1}{\rho} \log \mathbb{E}[\exp \{\rho \ell(f^\star_Z(Z)) \}] 
    = \Lambda_{Z}^{\star} (\rho, 0, \epsilon).
\end{align}
% %%%%%%

Next, we prove \eqref{Relation_CX_CZ}. For $D=0$, the guessing problem reduces to the lossless problem of Kuzuoka \cite{Kuzuoka}. Sakai and Tan \cite{Sakai} investigated the optimal guessing strategy for this setting, which is a special case of the lossy setting we consider. Let $G^\star_Z$ be the guessing function of the optimal guessing strategy for $Z$ with $D = 0$. From \cite[Lemma 10]{Sakai}, $G^\star_Z$ is given by
\begin{align}
    G^\star_Z(i) &= i, \quad (i=1,2, \ldots, l^*-1), \\
    G^\star_Z(l^*) &=     
    \begin{cases}
        0, & \mathrm{with ~ probability} ~ \alpha, \\
        l^*, &\mathrm{with ~ probability} ~ 1-\alpha, 
    \end{cases} \\
    G^\star_Z(j) &= 0, \quad \left(j=l^*+1, \ldots, \left \lceil \frac{|\mathcal{X}|}{L} \right \rceil \right),
\end{align}
where $\alpha$ is chosen so that \eqref{AppendixE_choice_alpha} holds. Noting that $l^* = i_Z^*$, we have 
\begin{align}
    M_{Z}^{\star} (\rho, 0, \epsilon)
    &=\mathbb{E}[G^\star_Z(Z)^\rho] \label{OptimalGuessinMoment_Z_1} \\
    &=\sum_{i=1}^{i_Z^*} Q^\epsilon_Z(i) i^\rho. \label{OptimalGuessinMoment_Z_2}
\end{align}
Combining \eqref{AchievabilityStep(c)} and \eqref{OptimalGuessinMoment_Z_2}, we obtain \eqref{Relation_CX_CZ}. 
%%%%%%%

%%%%%%%
Finally, we prove \eqref{Relation_LambdaZ_CZ}. From the definitions of the foregoing $f^\star_Z$ and $G^\star_Z$, it holds that
\begin{align}
    \Lambda_{Z}^{\star} (\rho, 0, \epsilon)
    & = \frac{1}{\rho} \log \mathbb{E}[\exp (\rho \ell(f^\star_Z(Z)))] \\
    & = \frac{1}{\rho} \log \left( \sum_{z=1}^{l^* - 1} P_Z(z) \exp (\rho \ell(f^\star_Z(z))) + (1-\alpha) P_Z(l^*) \exp (\rho \ell({\sf s}^+)) + \alpha P_Z(l^*) \exp (\rho \ell(\varnothing)) \right. \nonumber \\
    & \qquad \qquad \qquad \left. + \sum_{z=l^* +1}^{\left \lceil \frac{|\mathcal{X}|}{L} \right \rceil} P_Z(z) \exp (\rho \ell(\varnothing)) \right) \\
    & = \frac{1}{\rho} \log \left( \sum_{z=1}^{l^* - 1} P_Z(z) \exp (\rho \ell(f^\star_Z(z))) + (1-\alpha) P_Z(l^*) \exp (\rho \ell({\sf s}^+)) + \alpha P_Z(l^*) + \sum_{z=l^* +1}^{\left \lceil \frac{|\mathcal{X}|}{L} \right \rceil} P_Z(z) \right) \label{AppendixE_Lambda_Upper_stepA} \\
    & = \frac{1}{\rho} \log \left( \sum_{z=1}^{l^* - 1} P_Z(z) \exp (\rho \ell(f^\star_Z(z))) + (1-\alpha) P_Z(l^*) \exp (\rho \ell({\sf s}^+)) + \epsilon \right) \label{AppendixE_Lambda_Upper_stepB} \\
    & = \frac{1}{\rho} \log \left( \sum_{z=1}^{l^* - 1} P_Z(z) \exp (\rho \lfloor \log G^\star_Z(z) \rfloor) + (1-\alpha) P_Z(l^*) \exp (\rho \lfloor \log l^* \rfloor) + \epsilon \right) \label{AppendixE_Lambda_Upper_stepC} \\
    & \leq \frac{1}{\rho} \log \left( \sum_{z=1}^{l^* - 1} P_Z(z) \exp (\rho \log G^\star_Z(z) ) + (1-\alpha) P_Z(l^*) \exp (\rho \log l^*) + \epsilon \right) \label{AppendixE_Lambda_Upper_stepD} \\
    & = \frac{1}{\rho} \log \left( \sum_{z=1}^{l^* - 1} P_Z(z) G^\star_Z(z)^\rho + (1-\alpha) P_Z(l^*) (l^*)^\rho + \epsilon \right) \\
    & = \frac{1}{\rho} \log \left( \mathbb{E}[G^\star_Z(Z)^\rho] + \epsilon \right) \\
    & = \frac{1}{\rho} \log \left( M_{Z}^{\star} (\rho, 0, \epsilon) + \epsilon \right), 
\end{align}
where \eqref{AppendixE_Lambda_Upper_stepA} is due to $\ell(\varnothing)=0$; \eqref{AppendixE_Lambda_Upper_stepB} follows from \eqref{AppendixE_choice_alpha}; \eqref{AppendixE_Lambda_Upper_stepC} follows because $\ell(f^\star_Z(z)) = \lfloor \log G^\star_Z(z) \rfloor$ for $z=1, \ldots, l^* - 1$ and $\ell({\sf s}^+) = \lfloor \log l^* \rfloor$ from the definitions of $f^\star_Z$ and $G^\star_Z$; and \eqref{AppendixE_Lambda_Upper_stepD} is due to $\lfloor a \rfloor \leq a$ for $a \in \mathbb{R}$.
%%%%%%%
Moreover, we have
\begin{align}
    \Lambda_{Z}^{\star} (\rho, 0, \epsilon)
    & > \frac{1}{\rho} \log \left( \sum_{z=1}^{l^* - 1} P_Z(z) \exp (\rho \log G^\star_Z(z) - \rho) + (1-\alpha) P_Z(l^*) \exp (\rho \log l^* - \rho) + \epsilon \right) \label{AppendixE_Lambda_Lower_stepA} \\
    & = \frac{1}{\rho} \log \left( 2^{-\rho} \left[ \sum_{z=1}^{l^* - 1} P_Z(z) G^\star_Z(z)^\rho + (1-\alpha) P_Z(l^*) (l^*)^\rho \right] + \epsilon \right) \\
    & = \frac{1}{\rho} \log \left( 2^{-\rho} \mathbb{E}[G^\star_Z(Z)^\rho] + \epsilon \right)
    = \frac{1}{\rho} \log \left( 2^{-\rho}  M_{Z}^{\star} (\rho, 0, \epsilon) + \epsilon \right),
\end{align}
where \eqref{AppendixE_Lambda_Lower_stepA} follows from \eqref{AppendixE_Lambda_Upper_stepC} and $\lfloor a \rfloor > a - 1$ for $a \in \mathbb{R}$.
%%%%%%%
\section*{Acknowledgment}
H. Joudeh gratefully acknowledges many discussions with H. Wu on guessing, source coding, and log-loss.
%%%%%%%
\bibliographystyle{IEEEtran}
\bibliography{refs}
%\begin{IEEEbiographynophoto}{Shota Saito}received the B.E., M.E., and Ph.D. degrees in applied mathematics from Waseda University, Tokyo, Japan, in 2013, 2015, and 2018, respectively. From 2018 to 2021, he was an Assistant Professor at the Department of Applied Mathematics, Waseda University, Tokyo, Japan. Since 2021, he has been an Associate Professor with the Faculty of Informatics, Gunma University, Gunma, Japan. His research interests include information theory and its applications for machine learning and information-theoretic security. He is a recipient of IEEE IT Society Japan Chapter Young Researcher Best Paper Award and Student Paper Award in the 2016 International Symposium on Information Theory and Its Applications. He also received Waseda University Azusa Ono Memorial Award (Academic) in 2016 and Symposium on Information Theory and its Applications Young Researcher Paper Award in 2018. He is a member of the Institute of Electronics, Information and Communication Engineers (IEICE) and the Institute of Electrical and Electronics Engineers (IEEE).
%\end{IEEEbiographynophoto}

%\begin{IEEEbiographynophoto}{Hamdi Joudeh}
%\end{IEEEbiographynophoto}
\end{document}